\documentclass[aps, prd, superscriptaddress, preprintnumbers, showpacs]{revtex4}
\usepackage[english]{babel}
\usepackage{graphicx,amsfonts,amsmath,amssymb,amstext}
\usepackage{float,wrapfig}
\usepackage{subfigure, psfrag}
\usepackage{dsfont}
\usepackage{epsfig}
\usepackage{color}
\usepackage{bm}
\usepackage{multirow,textcomp}
\usepackage{natbib}
\usepackage[normalem]{ulem}
\usepackage{hyperref}

\newcommand{\be}{\begin{equation}}
\newcommand{\ee}{\end{equation}}
\newcommand{\ba}{\begin{eqnarray}}
\newcommand{\ea}{\end{eqnarray}}

\definecolor{purple}{rgb}{0.8,0,0.6}

\makeatletter
\newcommand{\vast}{\bBigg@{2}}
\newcommand{\Vast}{\bBigg@{3}}
\makeatother

\begin{document}
\title{Influence of tachyonic instability on the Schwinger effect in Higgs inflation model}
\date{\today}

\author{Mehran  Kamarpour}
\affiliation{Physics Faculty, Taras Shevchenko National University of Kyiv, 64/13, Volodymyrska str., 01601 Kyiv, Ukraine. Email: mehrankamarpour@yahoo.com}

\begin{abstract}
	We investigate the  influence of the tachyonic instability on the  Schwinger effect in Higgs inflation model.In this work we  identify the standard horizon scale $ k_{H}=aH $ and    the tachyonic instability  $ k_{H}=aH|\zeta| ,  \zeta=\frac{{I}^{\prime}\left(\phi\right)\dot{\phi}}{H}  $.This is the horizon scale in which the given Fourier begins to  become tachyonically unstable.Influence of  this scale appears by vanishing electromagnetic field energy density and energy density of created charged particles due to the Schwinger effect at the very beginning of inflation but does not alter conclusions of our previous work in Refs.\cite{Kamarpour:2022,Kamarpour:2023-I}.We use two coupling functions to break conformal invariance of maxwell action.The simplest coupling function $ I\left(\phi\right)=\chi_{1}\frac{\phi}{M_{p}} $  and a curvature based  coupling function $ I\left(\phi\right)= 12\chi_{1}e^{\left(\sqrt{\frac{2}{3}}\frac{\phi}{M_{p}}\right)}\left[\frac{1}{3M_{p}^{2}}\left(4V\left(\phi\right)\right)+\frac{\sqrt{2}}{\sqrt{3}M_{p}}\left(\frac{dV}{d\phi}\right)\right] $ where $V\left(\phi\right)  $ is the potential of Higgs inflation in Refs.\cite{Kamarpour:2022,Kamarpour:2023-I}.In fact, we find that only at the very beginning of inflation both energy densities of electromagnetic field and created charged particles vanish due to effect of  tachyoinc instability.

\end{abstract}

\pacs{000.111\\
	Keywords:magnetogenesis,Axial coupling coupling, Higgs inflation,The Schwinger effect}

\maketitle
%\graphicspath{{Figures/}}

\tableofcontents

\section{Introduction}
\label{sec-intro}
We have recently shown that  in Higss inflation model , the Schwinger effect with conformal coupling or non-coupling function to gravity is quite negligible whereas at the same model with the simplest coupling function is considerable and even in latter work shows the so-called the Schwinger reheating scenario-framework.See \cite{Kamarpour:2022,Kamarpour:2023-I}.In both works we identified $k_{H}=aH $ as the standard horizon scale for boundary terms.

In this paper we identify $ k_{H}=aH |\zeta| ,      \zeta=\frac{{I}^{\prime}\left(\phi\right)\dot{\phi}}{H} $ as horizon scale in which the given Fourier begins to become tachyoically unstable.We should set the first and second derivatives of vector potential equal to zero in order to find tachyoinc instability.

In order to show whether the Schwinger effect exists or does not exist we need to add some boundary terms in required equation.These boundary terms are quiet different in kinetic couplings  and axial couplings.

Previously , Refs.\cite{Kronberg:1994,Grasso:2001,Widrow:2002,Giovannini:2004,Kandus:2011,Durrer:2013,Subramanian:2016} have mentioned and detailed about magneto-genesis and also cosmic microwave background observations put upper and lower bounds to the strength of primordial magnetic field\cite{Planck:2015,Planck:2018,Sutton:2017,Jedamzik:2018}.In addition, gamma rays emitted by distant blazars  have shown the strength of primordial magnetic fields  $ B_{0} $ ranging from  $ 10^{-17} $ to $ 10^{-9} $ gauss\cite{Neronov:2010,Tavecchio:2010,Taylor:2011,Caprini:2015}.Origins of these magnetic field previously have been studied by several authors \cite{Biermann:1950,Zeldovich:1980book,Lesch:1995,Kulsrud:1997,Colgate:2001,Rees:1987,Daly:1990,Ensslin:1997,Bertone:2006}.

However, one may find very useful discussions in Refs.\cite{Turner:1988,Hogan:1983,Quashnock:1989,Vachaspati:1991} about the primordial nature of detected magnetic fields as well as the most natural mechanism i.e. inflation which is a period of rapid expansion in early Universe , for the generation of large-coherence-scale magnetic fields \cite{Turner:1988}.

It is said that during inflation, quantum fluctuations of massless scalar and tensor fields can be amplified significantly which is source of  the large-scale structures observed in Universe now-days.\cite{Mukhanov:1981,Hawking:1982,Starobinsky:1982,Guth:1982,Bardeen:1983}.Additionally, it is indicated that this amplification plays main role for the generation of relic gravitational waves\cite{Grishchuk:1975,Starobinsky:1979,Rubakov:1982}.

However, the conformal invariance of the Maxwell action should be broken because it does not allow of generation of large-scale magnetic fields\cite{Parker:1968}.Therefore, in this paper we allow the conformal invariance to be broken by axial coupling.We have already shown in Higgs inflation model kinetic coupling $ I^{2}\left(\phi\right)F_{\mu\nu}F^{\mu\nu} $ which first introduced by Ratra  \cite{Ratra:1992} and   discussed in Refs.\cite{Giovannini:2001,Bamba:2004,Martin:2008,Demozzi:2009,Kanno:2009,Ferreira:2013,Ferreira:2014,Vilchinskii:2017}. is not successful in order to produce magnetic field whereas in  axial coupling  by interaction term of the form $ I\left(\phi\right)F_{\mu\nu}\tilde{F}^{\mu\nu} $ \cite{Dolgov:1993,Gasperini:1995,Giovannini:2000,Atmjeet:2014} where $ I\left(\phi\right) $ is  a coupling function of the inflaton field  $ \phi $ and $ F_{\mu\nu}=\partial_{\mu} A_{\nu}-\partial_{\nu}A_{\mu} $ is the electromagnetic field tensor and $ \tilde{F}^{\mu\nu}=\frac{1}{2}\frac{\epsilon^{\mu\nu\rho\sigma}}{\sqrt{-g}}F_{\rho\sigma} $ we can produce magnetic field\cite{Kamarpour:2021}.

Additionally, in axial coupling the electric energy density is almost equal to the magnetic energy density , i.e. $ \rho_{E}\sim \rho_{B} $ \cite{Figueroa:2018,Notari:2016,Fujita:2015,Kamarpour:2021}.

The Schwinger effect first considered by Schwinger \cite{Schwinger:1951} and it is the process of producing charged particles from vacuum by applying strong electric field.

When inflaton filed coupled to electromagnetic field ,strong electric field can be generated and accordingly charged particles can be produced from vacuum.The case of a constant and homogeneous electric field in  de~Sitter  space-time can be found in Refs. \cite{Afshordi:2014,Froeb:2014,Bavarsad:2016,Stahl:2016a, Stahl:2016b,Hayashinaka:2016a,Hayashinaka:2016b,Sharma:2017, Tangarife:2017,  Hayashinaka:2018, Hayashinaka:thesis, Stahl:2018, Geng:2018, Bavarsad:2018}.

We should emphasis that in expanding Universe the Schwinger effect has some specific effects with some expressions for produced charged particles by strong electric field.These features of cosmological Schwinger effect for very small mass or massless particles  have been studied in Refs.\cite{Afshordi:2014,Hayashinaka:2016a,Hayashinaka:2016b,Hayashinaka:2018,Hayashinaka:thesis,Stahl:2018}. 

It should be noted that we must use expressions for the Schwinger effect only in case of a  time-dependent electric field in the strong-field regime \cite{Kitamoto:2018} because the case of constant and homogeneous electric field contradicts the second law of thermodynamics by  requiring  the existence of ad hoc currents \cite{Giovannini:2018a}.    

It has been shown that the Schwinger effect in weak field regime is negligible whereas in strong field regime is considerable\cite{Sobol-Gorbar:2021}.Therefore, in this paper we only consider the strong field regime.

In this paper, we investigate the Schwinger effect by axial coupling in Higgs  inflation model with two coupling functions.  The paper is organized as follows: we determine  the model and find the solution for the background equations in the Higgs inflation model in Sect.~\ref{sec-inflation}, where we also consider the axial coupling of the inflation field to the electromagnetic field with two coupling functions. We then obtain the mode-function and estimate the range of parameters for which the back-reaction problem does not occur for our model. In this part , we consider a system of self-consistent equations, including the Schwinger effect. Section~\ref{sec-Schwinger} discusses the main idea of the Schwinger effect, including the Schwinger source term. In Sect.~\ref{sec-Numerical}, we perform numerical calculations for both scenarios, when we use the standard  horizon scale $ k_{H}=aH $  and  with the scale  $ k_{H}=aH|\zeta| ,  \zeta=\frac{{I}^{\prime}\left(\phi\right)\dot{\phi}}{H}  $ at which a given Fourier begins to become tachyonically unstable  for two coupling functions and compare them. The summary of the obtained results is given in Sect.~\ref{sec-conclusion}.

\section{Higgs inflation}
\label{sec-inflation}
We use the potential of the Higgs inflation model that was introduced in Refs.~\cite{Kamarpour:2021,Kamarpour:G,Kamarpour:2022,Kamarpour:2023-I}. For more details see our previous works in Refs.~\cite{Kamarpour:2021,Kamarpour:G,Kamarpour:2022,Kamarpour:2023-I}.This potential is valid for all stage of inflation.  
\begin{equation}
\label{potential}
V\left(\phi\right)=\frac{\lambda}{4}\frac{M^{4}_{p}}{\xi^{2}_{h}}\left(1-e^{-\sqrt{\frac{2}{3}}\frac{\phi}{M_{p}}}\right)^{2}\left[1+\frac{\textbf{A}_{I}}{32\pi^{2}}\ln\left(e^{\sqrt{\frac{2}{3}}\frac{\phi}{M_{p}}}-1\right)\right],
\end{equation}
In above equation  $\textbf{A}_{I}=\textbf{A}-12\lambda $ is anomalous scaling and $ \lambda=0.12029 $ , $ \xi_{h} $ is non-minimal coupling constant.

We consider a spatially flat Friedmann--Lema\^{i}tre--Robertson--Walker(FLRW) Universe with metric tensor
\begin{equation}
\label{metric}
g_{\mu\nu}={\rm diag}\,(1,\,-a^{2},\,-a^{2},\,-a^{2}), \quad \sqrt{-g}=a^{3},
\end{equation}
and use the natural system of units where $\hbar=c=1$, $M_{p}=(8\pi G)^{-1/2}=2.4\cdot 10^{18}\,{\rm GeV}$ is a reduced Planck mass , and $ e=\sqrt{4\pi\alpha}\approx 0.3 $ is the absolute value of the electron's charge.

\subsection{Action}
The action of the inflaton field interacting with electromagnetic field  and electromagnetic field interacting with charged field $ \chi $ by axial coupling is given by
\begin{equation}
\label{action-1}
S=\int d^{4}x \sqrt{-g}\left[\frac{1}{2}\partial^{\mu}\phi\partial_{\mu}\phi-V(\phi)-\frac{1}{4}F_{\mu\nu}F^{\mu\nu}+\frac{1}{4}I\left(\phi\right)F_{\mu\nu}\tilde{F}^{\mu\nu}+L_{charged}\left(A,\chi\right)\right]
\end{equation}
In above equation , $ V\left(\phi\right) $ is the Higgs potential, $ I\left(\phi\right) $ is axial coupling function and $ L_{charged}\left(A,\chi\right) $ is a gauge-invariant Lagrangian of the charged field $ \chi $ (bosonic or fermionic) interacting with electromagnetic field.

Variation for inflaton field $\phi $ reads
\begin{equation}
\label{motion}
\frac{1}{\sqrt{-g}}\partial{_{\mu}}\left[\sqrt{-g}\partial{^{\mu}\phi}\right]+\frac{dV}{d\phi}
=\frac{1}{4}\frac{dI\left(\phi\right)}{d\phi}F_{\mu\nu}\tilde{F}^{\mu\nu}
\end{equation}
Also variation of gauge field $ A_{\mu} $ gives following relation
\begin{equation}
\label{Maxwell-Tilde}
\frac{1}{\sqrt{-g}}\partial_{\mu}\left[\sqrt{-g}I(\phi)\tilde{F}^{\mu\nu}-\sqrt{-g}F^{\mu\nu}\right]=-J^{\nu}
\end{equation}
In above equation $ \tilde{F}^{\mu\nu}=\frac{1}{2}\frac{\epsilon^{\mu\nu\rho\sigma}}{\sqrt{-g}}F_{\rho\sigma} $ in which $ \epsilon^{\mu\nu\rho\sigma} $ is the totally antisymmetric Levi-Civita symbol with $ \epsilon^{0123}=1 $.

By manipulation of equation (\ref{Maxwell-Tilde}) one finds following equations.
%Further manipulations of equation(\ref{Maxwell-Tilde}) gives following useful equation
\begin{equation}
\label{Maxwell-}
\frac{1}{\sqrt{-g}}\partial_{\mu}\left[\sqrt{-g}{F}^{\mu\nu}\right]-I^{\prime}\left(\phi\right)\partial_{\mu}\phi \tilde{ F}^{\mu\nu}=-J^{\nu}
\end{equation}
where $ I^{\prime}\left(\phi\right)=\frac{dI\left(\phi\right)}{d\phi} $.In above equation $ J^{\nu}$ is given by
\begin{equation}
J^{\mu}=\frac{\partial L_{charged}\left(A,\chi\right)}{\partial A_{\mu}}
\end{equation}

Another useful equation is given by following relation
\begin{equation}
\label{Maxwell-tilde}
\frac{1}{\sqrt{-g}}\partial_{\mu}\left[\sqrt{-g}\tilde{F}^{\mu\nu}\right]=0
\end{equation}
In Eq. (\ref{motion}) $ F_{\mu\nu}\tilde{F}^{\mu\nu}=-4 \textbf{E}\cdot\textbf{B} $ , by using this equation we obtain
\begin{equation}
\label{Back-reaction}
\ddot{\phi}+3H\dot{\phi}+\frac{dV\left(\phi\right)}{d\phi}=-I^{\prime}\left(\phi\right)\textbf{E}\cdot\textbf{B}
\end{equation} 

Note that we assume homogeneous  inflaton field $ \phi=\phi\left(t\right) $.We use Coulomb gauge for electromagnetic field , i.e. $ A_{\mu}=\left(0,\textbf{A}\right) $ and $ \nabla\cdot\textbf{A}=0    $ .

Therefore, the   electric and magnetic fields are given by following relations
\begin{equation}
\label{Electric-Magnetic}
\textbf{E}=-\frac{1}{a}\dot{\textbf{A}},\hspace{.5cm}\textbf{B}=\frac{1}{a^{2}}\nabla\times\textbf{A}
\end{equation}
In equation \ref{Electric-Magnetic} , $ a=a\left(t\right) $ is scale factor of FLRW Universe.In terms of electromagnetic field tensor $ F^{\mu\nu}=\partial^{\mu}A^{\nu}-\partial^{\nu}A^{\mu} $ and its dual  tensor $ \tilde{F}^{\mu\nu} $ the components of electric and magnetic fields are given by following relations
\begin{equation}
\label{EM-tensor}
F^{0i}=\frac{1}{a}E^{i}\hspace{.5cm},F_{ij}=a^{2}\epsilon_{ijk}B^{k}\hspace{.5cm},\tilde{F}^{0i}=\frac{1}{a}B^{i},\hspace{.5cm}\tilde{F}_{ij}=-a^{2}\epsilon_{ijk}E^{k}
\end{equation}
Note that $\epsilon_{ijk} $ is three dimensional Levi-Civita symbol and $ i,j,k  $ indicate components of 3-vectors. By using the components of electric and magnetic fields , i.e.Eqs. (\ref{EM-tensor}) and Eqs.(\ref{Maxwell-} ,  \ref{Maxwell-tilde}) we obtain following  system of closed equations.
\begin{equation}
\label{Electric-Current}
\dot{\textbf{E}}+2H\textbf{E}-\frac{1}{a}\nabla\times\textbf{B}-I^{\prime}\left(\phi\right)\dot{\phi}\textbf{B}=-a\textbf{J},
\end{equation} 
\begin{equation}
\label{Magnetic}
\dot{\textbf{B}}+2H\textbf{B}+\frac{1}{a}\nabla\times\textbf{E}=0,
\end{equation}
\begin{equation}
\nabla\cdot\textbf{B}=0,\hspace{.5cm}\nabla\cdot\textbf{E}=0
\end{equation}
It is more convenient to write  current $ \textbf{J} $ of charged particles in equation  (\ref{Electric-Current})  in terms of the generalized conductivity $ \sigma $ , so we have
\begin{equation}
\label{Current-Counductivty}
\textbf{J}=\frac{1}{a}\sigma\textbf{E}
\end{equation}

Let us look at equations (\ref{Electric-Current},\ref{Magnetic}).We have to write these equations in terms of energy densities for numerical calculations.Also, we should use approximation for electric and magnetic energy densities   in oder to close system of equations.Thus,  to obtain required relations for electric and magnetic energy densities it is more convenient to introduce energy-momentum tensor.
\begin{equation}
\label{Energy-Momentum}
T_{\mu\nu}=\frac{2}{\sqrt{-g}}\frac{\delta S}{\delta g^{\mu\nu}}=\partial_{\mu}\phi\partial_{\nu}\phi-g^{\alpha\beta}F_{\mu\alpha}F_{\nu\beta}-g_{\mu\nu}L_{0}+T^{charged}_{\mu\nu}
\end{equation} 
In above equation we use  $ L_{0}= \frac{1}{2}\partial^{\mu}{\phi}\partial_{\mu}\phi-V\left(\phi\right)-\frac{1}{4}F_{\mu\nu}F^{\mu\nu} $ .We know $ F_{\mu\nu}\tilde{F}^{\mu\nu}  $ does not appear in  the energy-momentum relation because it does not depend on metric.One may find energy density from $ T_{00} $ component.
\begin{equation}
\label{Total-Density}
\rho=\frac{1}{2}\dot{\phi}^{2}+V\left(\phi\right)+\frac{1}{2}\left(E^{2}+B^{2}\right)+\rho_{\chi}=\rho_{inf}+\rho_{EM}+\rho_{\chi}
\end{equation}
In equation (\ref{Total-Density}) $ \rho_{EM} $ is the energy density of electromagnetic filed and $ \rho_{\chi} $ is  the energy density of produced charged particles due to the Schwinger effect.Therefore, the Friedmann equation can be written by
\begin{equation}
\label{Friedmann-2}
H^{2}=\frac{1}{3M_{p}^{2}}\left(\rho_{inf}+\rho_{EM}+\rho_{\chi}\right)
\end{equation} 
Let us look at equation(\ref{Back-reaction}).In this equation the right hand side shows the  back-reaction term and it demonstrates helical nature of electromagnetic field( due to axial coupling ).In addition ,this term should be approximated by simpler relation such as $ \rho_{EM} $ in order to obtain system of closed equation for numerical calculations.We will return to this equation in section of numerical calculations and discuss about it. 

Equations (\ref{Electric-Current} , \ref{Magnetic}) in terms of energy density can be written  in terms of energy densities.By using equations (\ref{Electric-Current} , \ref{Current-Counductivty}) and (\ref{Magnetic}) we write
\begin{equation}
\label{EM-Density}
\dot{\rho}_{EM}+4H\rho_{EM}+ 2\sigma\rho_{E}-I^{\prime}\left(\phi\right)\dot{\phi}\textbf{E}\cdot\textbf{B}=0
\end{equation}
In numerical calculations we show that in  equation (\ref{EM-Density}) the term $ 2\sigma\rho_{E} $ indicates dissipation of the electromagnetic energy density due to the Schwinger effect.We already discussed  about the term $ \textbf{E}\cdot\textbf{B} $.In fact  the term $ I^{\prime}\left(\phi\right)\dot{\phi}\textbf{E}\cdot\textbf{B} $ describes axial coupling nature of electromagnetic field and inflaton field and implies transfer of energy density from inflaton field to electromagnetic field.
\begin{figure}[ht]
	\centering
	\includegraphics[width=0.45\textwidth]{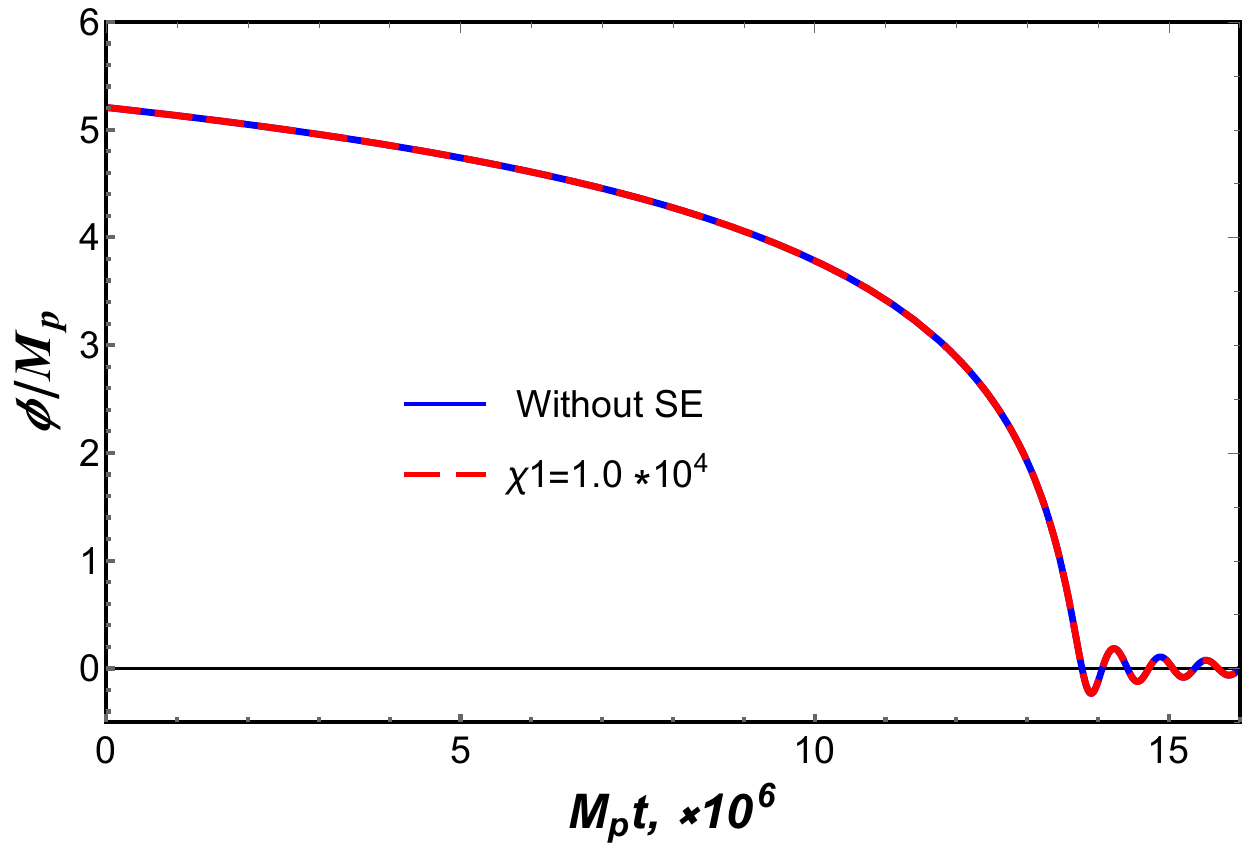}
	\hspace*{2mm}
	\includegraphics[width=0.45\textwidth]{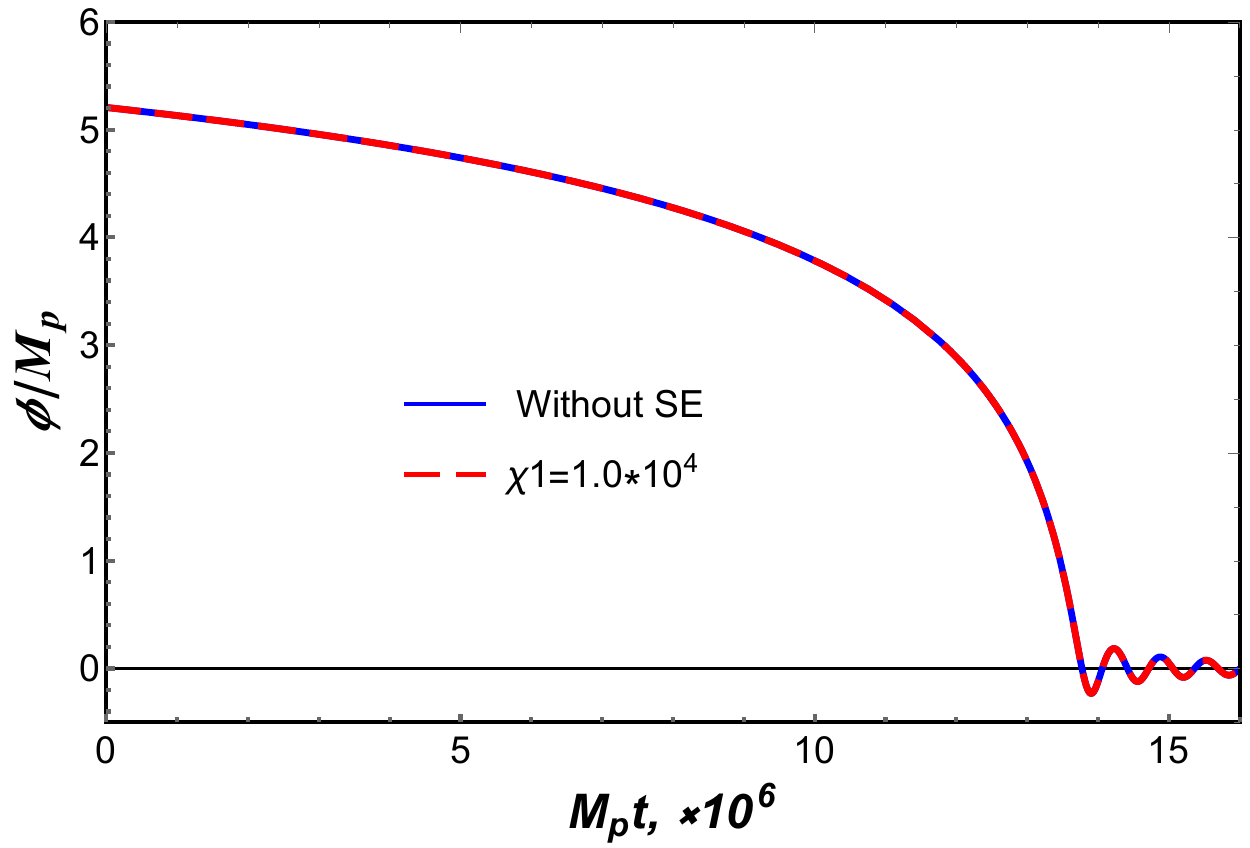}
	\caption{Panel a: using standard horizon scale $ k_{H}=aH $   and  panel b:  using effect of tachyonic instability by choosing $ K_{H}=aH|\zeta| $.The time dependence of  inflaton field for non-minimal coupling function to gravity   Eq.\ref{Coupling-3} (a) and  for non-minimal coupling to gravity  Eq.\ref{Coupling-3} (b)  for the same values of parameter $ \chi_{1} $.Both without the Schwinger effect and with the Schwinger effect.In both figures we see the Schwinger effect is quite negligible and using tachyonic instability has no effect on inflaton field.}
	\label{Inflaton-1}
\end{figure}
\begin{figure}[ht]
	\centering
	\includegraphics[width=0.45\textwidth]{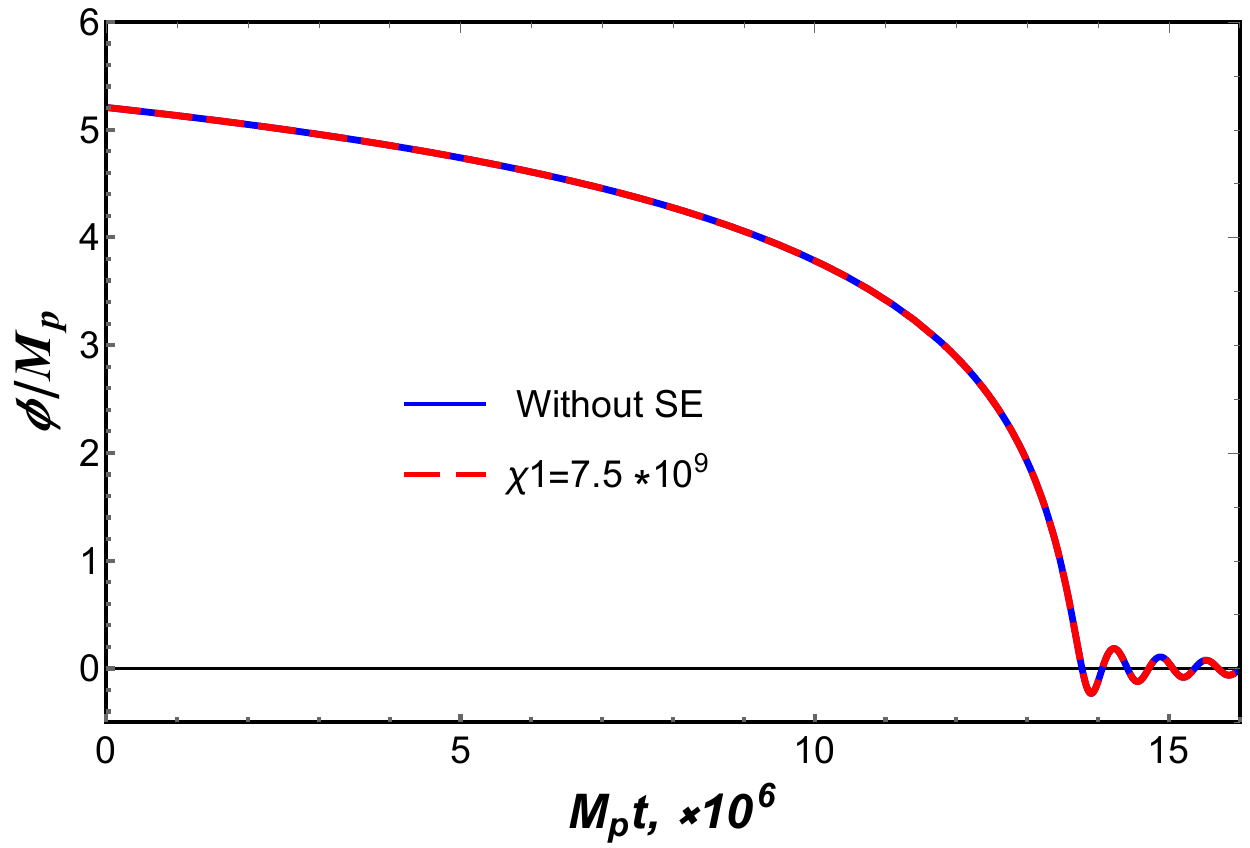}
	\hspace*{2mm}
	\includegraphics[width=0.45\textwidth]{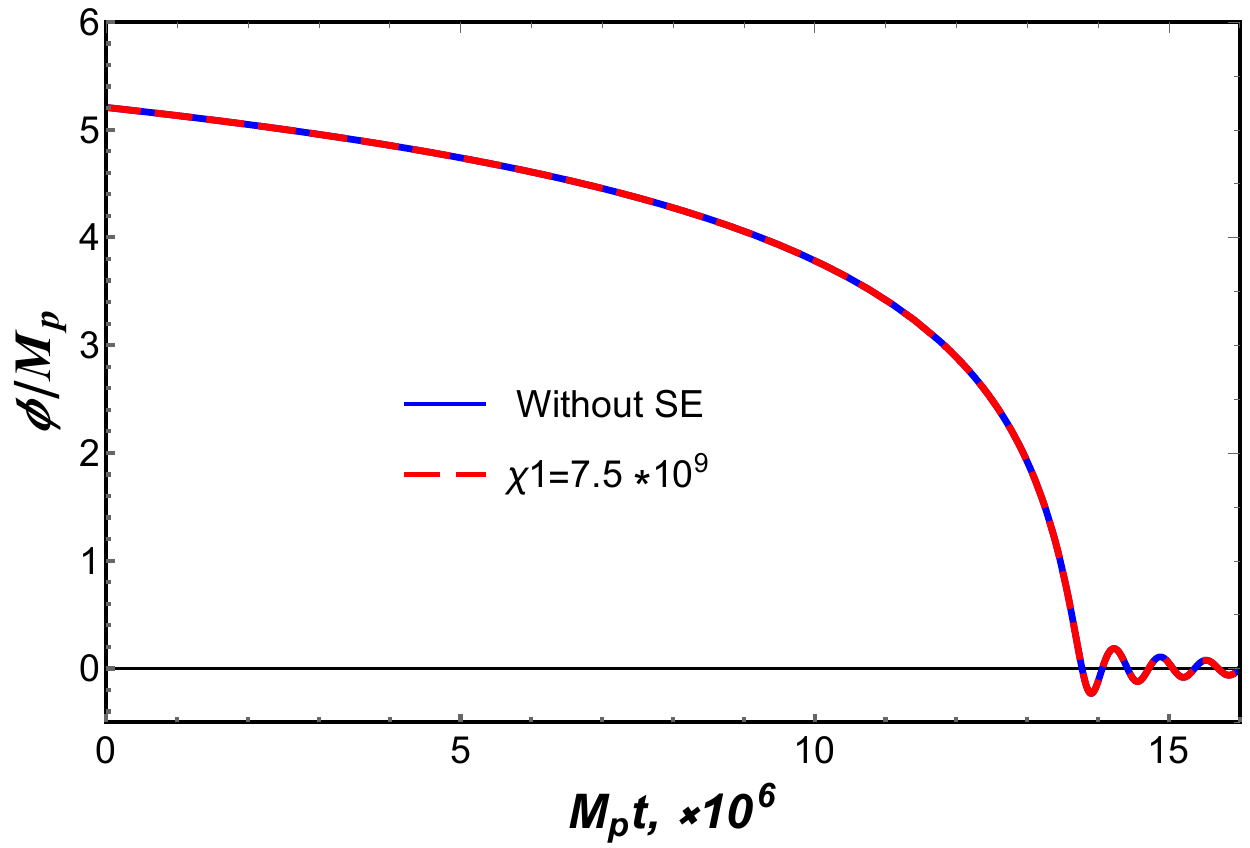}
	\caption{Panel a: using standard horizon scale $ k_{H}=aH $   and  panel b:  using effect of tachyonic instability by choosing $ K_{H}=aH|\zeta| $.The time dependence of  inflaton field for non-minimal coupling function to gravity   Eq.\ref{Coupling-3} (a) and  for non-minimal coupling to gravity  Eq.\ref{Coupling-3} (b)  for the same values of parameter $ \chi_{1} $.Both without the Schwinger effect and with the Schwinger effect.In this figure we increase the value of coupling parameter $ \chi_{1} $ again  but in both figures we see the Schwinger effect is quite negligible and using tachyonic instability has no effect on inflaton field.}
	\label{Inflaton-3}
\end{figure}

\begin{figure}[ht]
	\centering
	\includegraphics[width=0.45\textwidth]{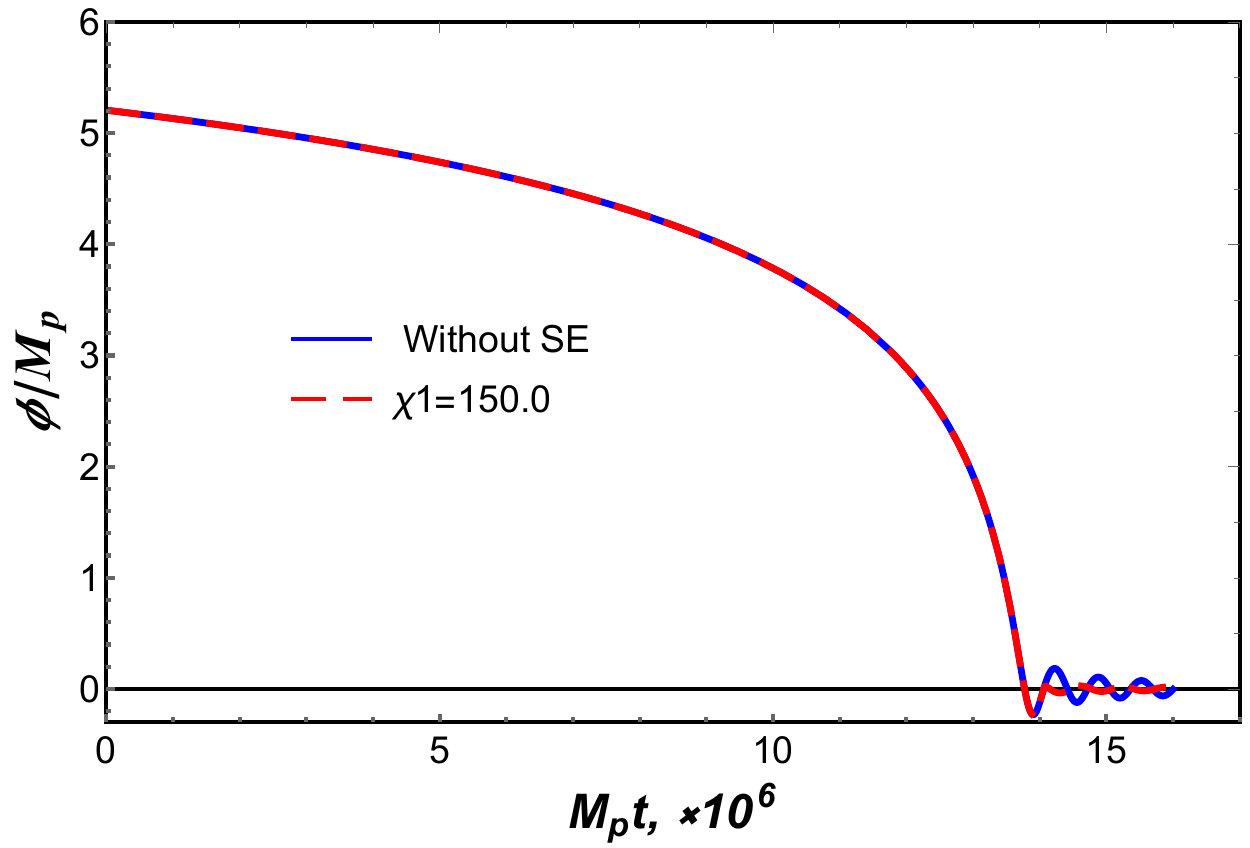}
	\hspace*{2mm}
	\includegraphics[width=0.45\textwidth]{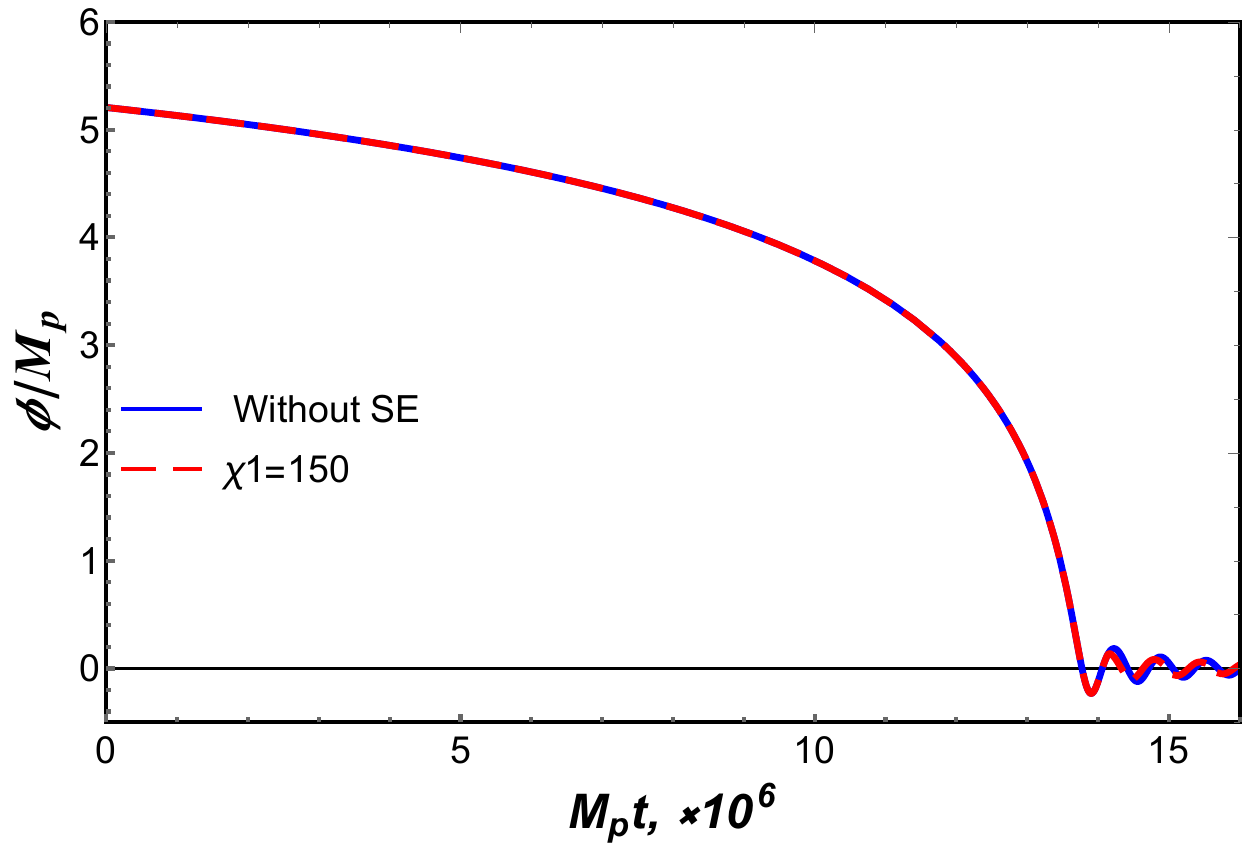}
	\caption{Panel a: using standard horizon scale $ k_{H}=aH $   and  panel b:  using effect of tachyonic instability by choosing $ K_{H}=aH|\zeta| $.The time dependence of  inflaton field for the simplest coupling function  Eq.\ref{Coupling-1} (a) and  for the same coupling  Eq.\ref{Coupling-1} (b)  for the same values  of parameter $ \chi_{1} $.Both without(blue) the Schwinger effect and with (red)the Schwinger effect.We see small differences in oscillatory parts of two panels.In fact choosing tachyonic instability  $ K_{H}=aH|\zeta| $ has effect on inflaton field in oscillatory part.}
	\label{Inflaton-4}
\end{figure}

\begin{figure}[ht]
	\centering
	\includegraphics[width=0.45\textwidth]{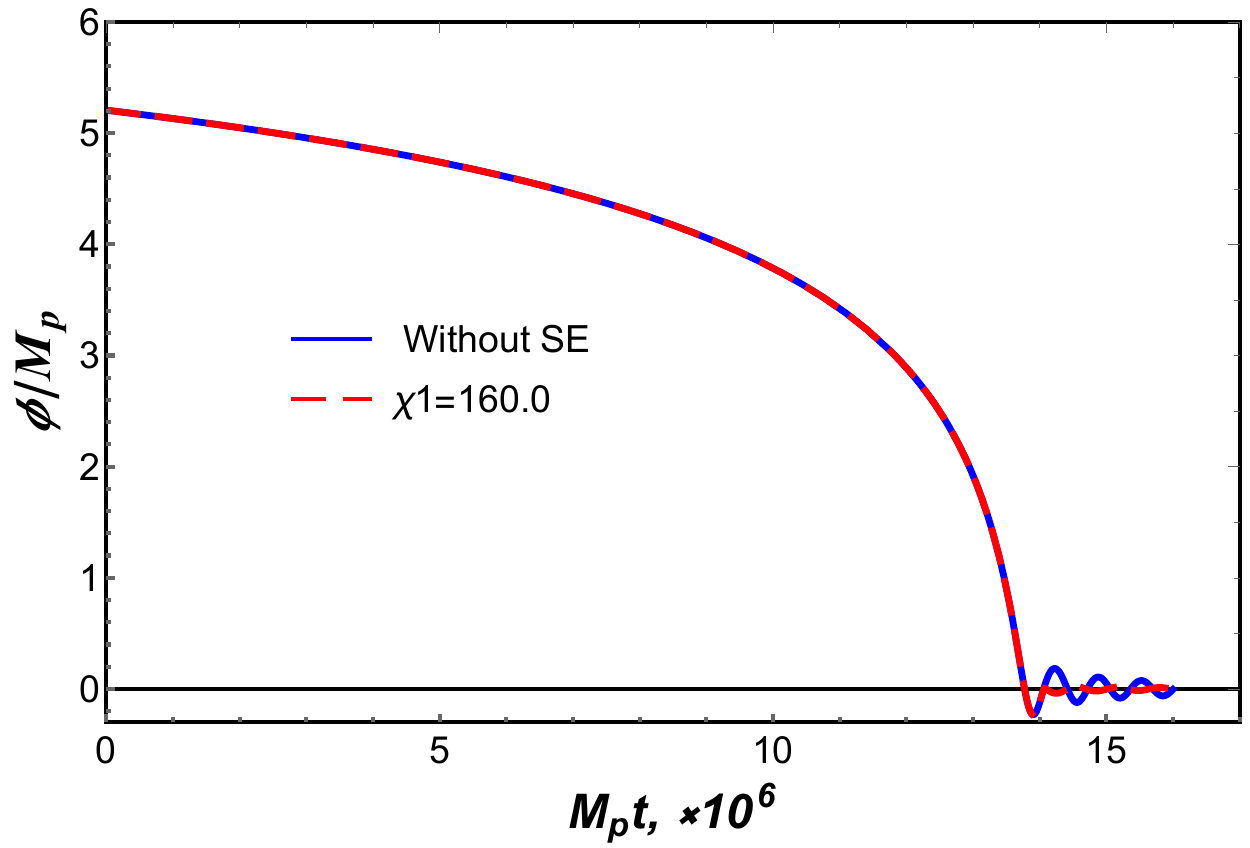}
	\hspace*{2mm}
	\includegraphics[width=0.45\textwidth]{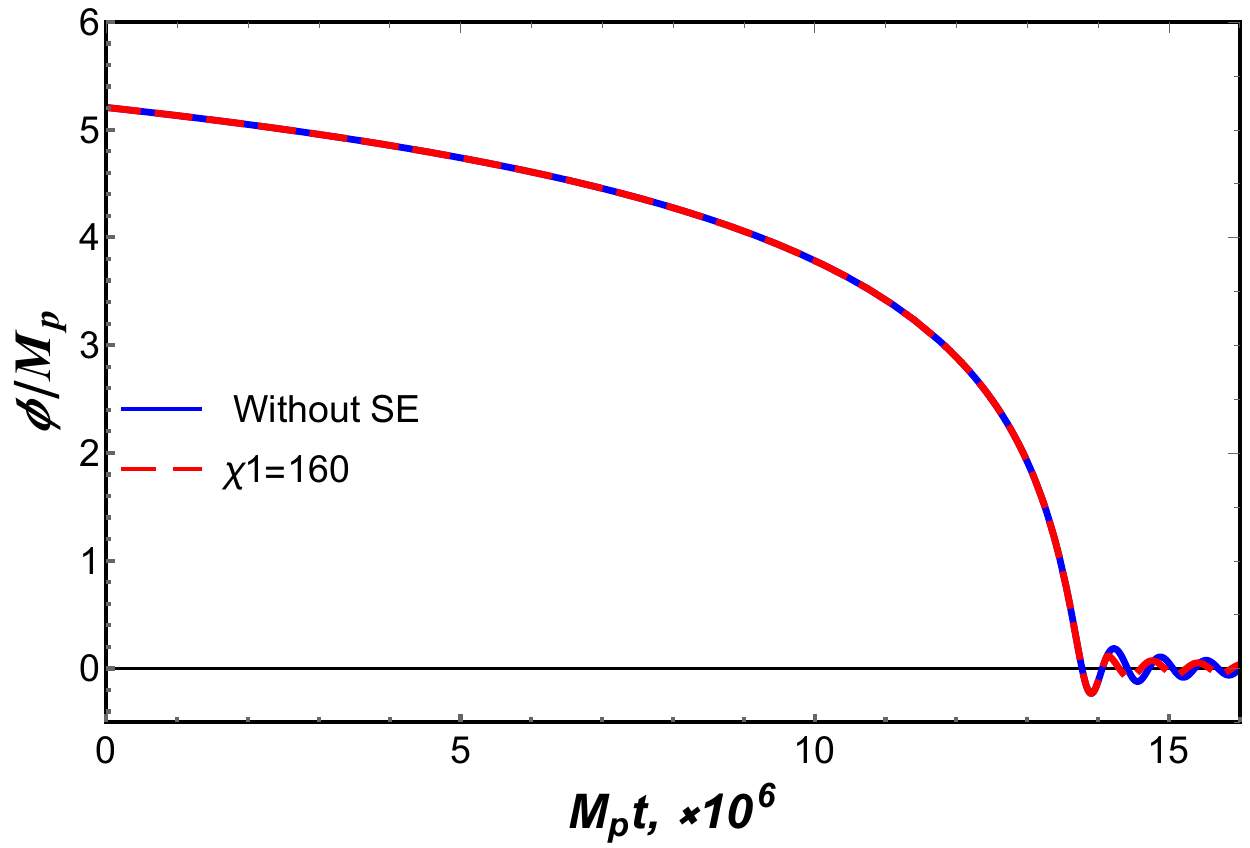}
	\caption{Panel a: using standard horizon scale $ k_{H}=aH $   and  panel b:  using effect of tachyonic instability by choosing $ K_{H}=aH|\zeta| $.The time dependence of  inflaton field for the simplest coupling function  Eq.\ref{Coupling-1} (a) and  for the same coupling  Eq.\ref{Coupling-1} (b)  for the same values  of parameter $ \chi_{1} $.Both without(blue) the Schwinger effect and with (red)the Schwinger effect.We see small differences in oscillatory parts of two panels.In fact , choosing tachyonic instability  $ K_{H}=aH|\zeta| $ has effect on inflaton field in oscillatory part.Note that we increase the value of coupling parameter $ \chi_{1} $ and we see  that the effect of tachyoinc instability  in panel (b) is bigger.}
	\label{Inflaton-5}
\end{figure}

\begin{figure}[ht]
	\centering
	\includegraphics[width=0.45\textwidth]{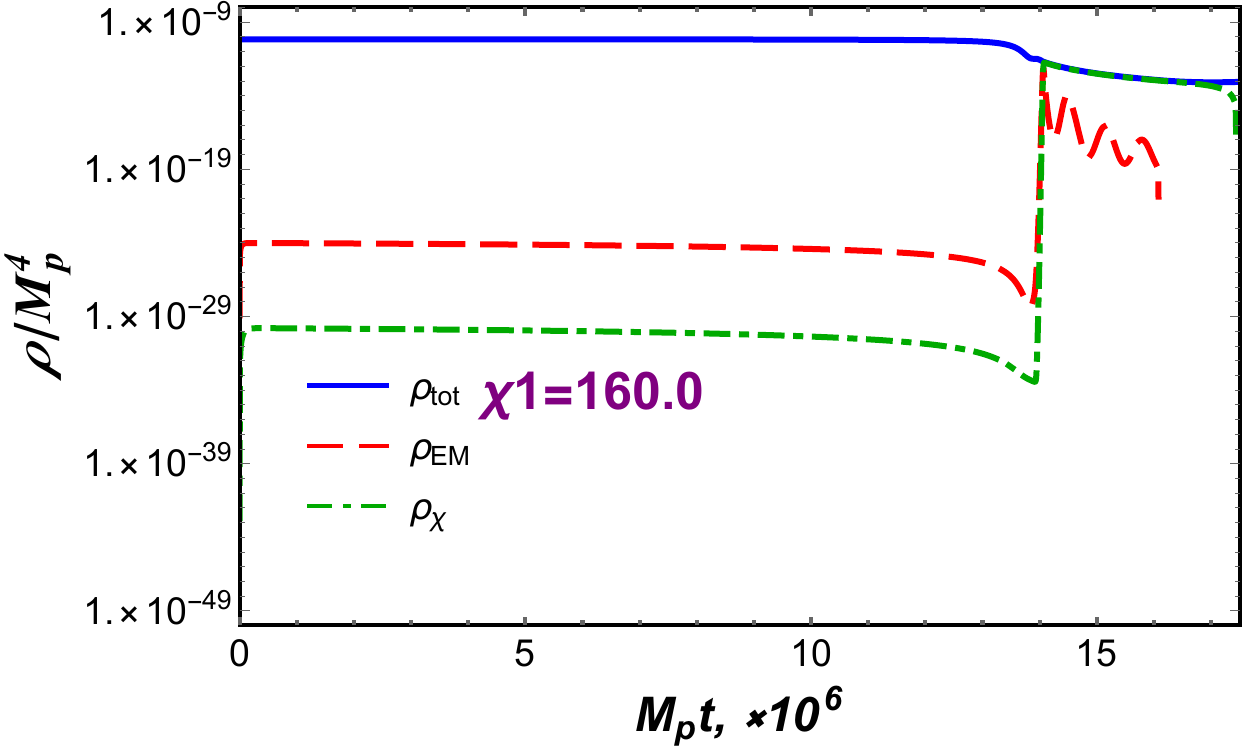}
	\hspace*{2mm}
	\includegraphics[width=0.45\textwidth]{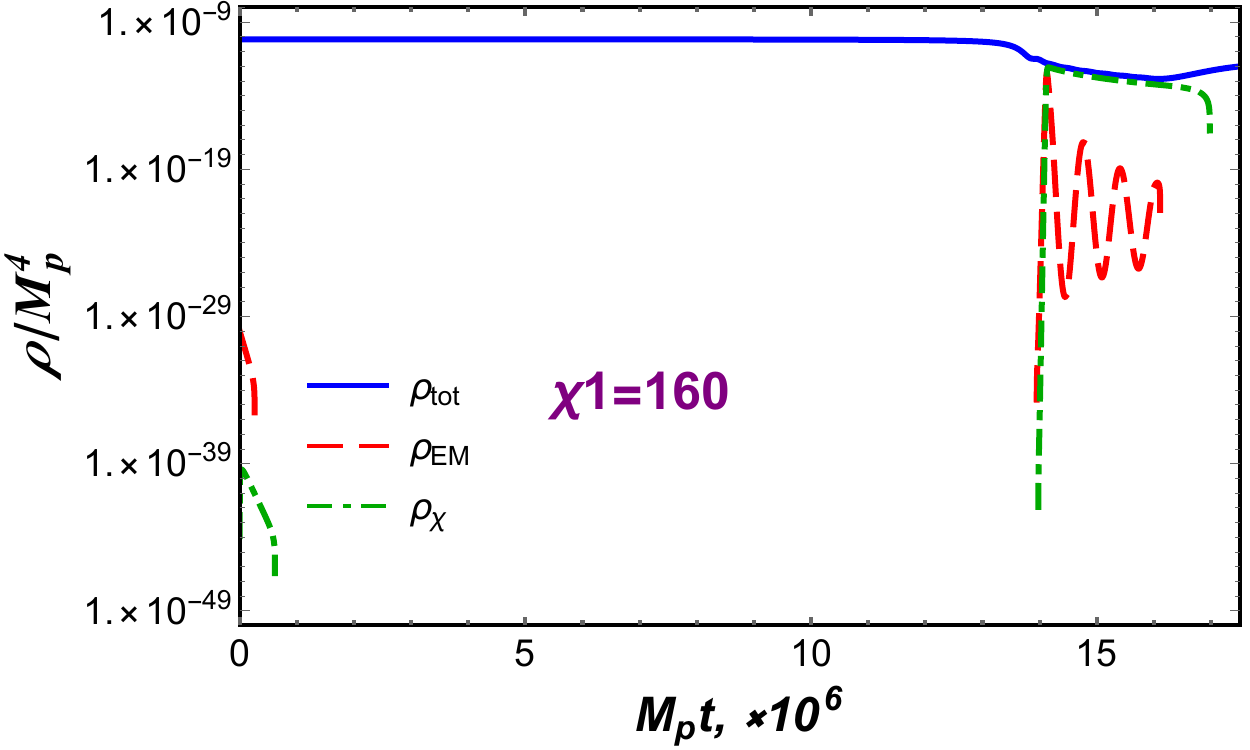}
	\caption{Panel a: using standard horizon scale $ k_{H}=aH $   and  panel b:  using effect of tachyonic instability by choosing $ K_{H}=aH|\zeta| $ .The time dependence (a)  of energy densities  for $ \chi_{1}=160 $ and  the simplest coupling function  Eq.\ref{Coupling-1} and  (b) for  the same coupling function  Eq.\ref{Coupling-1} and $ \chi_{1}=160 $ .In both panels there is no back-reaction problem . In each panel $ \rho_{tot} $ (blue), $ \rho_{EM}$ (red dashed line), $ \rho_{\chi} $ (green dashed line )  show the total energy density, electromagnetic energy density and energy density of charged particles due to  the Schwinger effect respectively.Note that in panel (a) we do not see effect of tachyonic instability because we use standard horizon scale  $ k_{H}=aH $ .In panel (b) we see that at very beginning of inflation both energy densities of electromagnetic field and produced charged particles due to the Schwinger effect vanish.This is the effect of tachyonic instability.}
	\label{Density-sim-1}
\end{figure}
\begin{figure}[ht]
		\centering
		\includegraphics[width=0.45\textwidth]{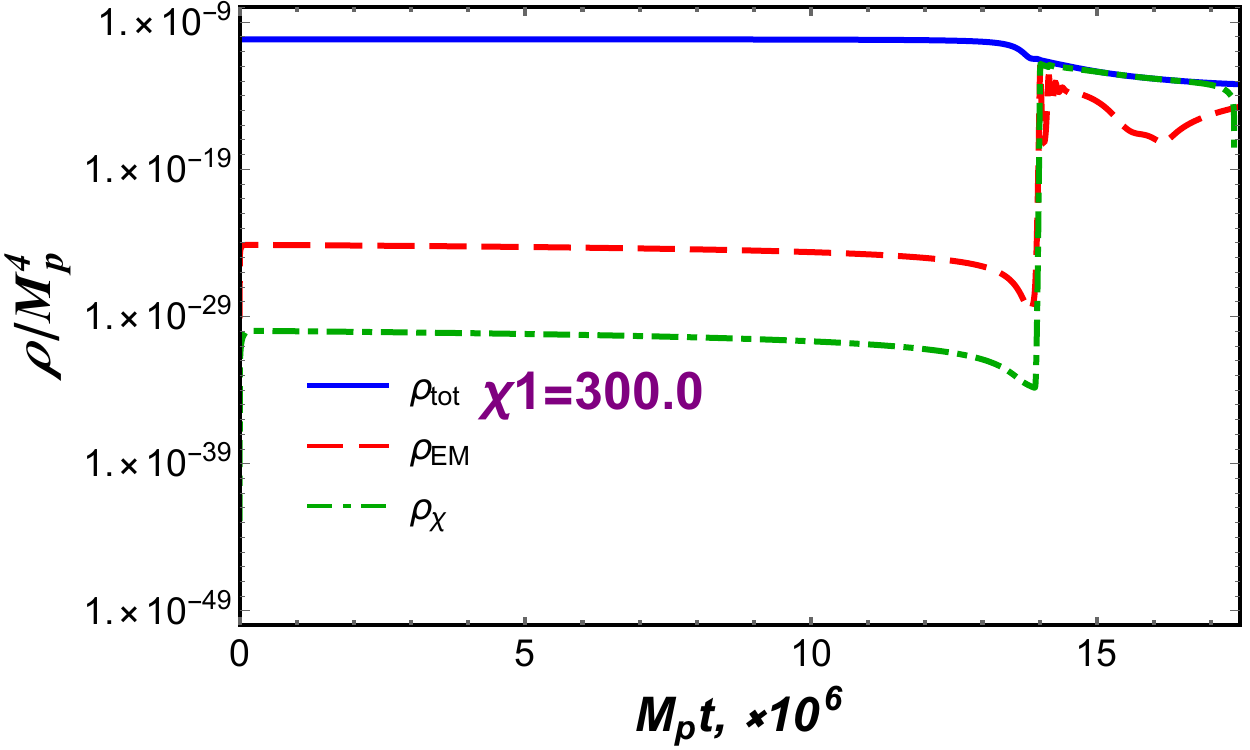}
		\hspace*{2mm}
		\includegraphics[width=0.45\textwidth]{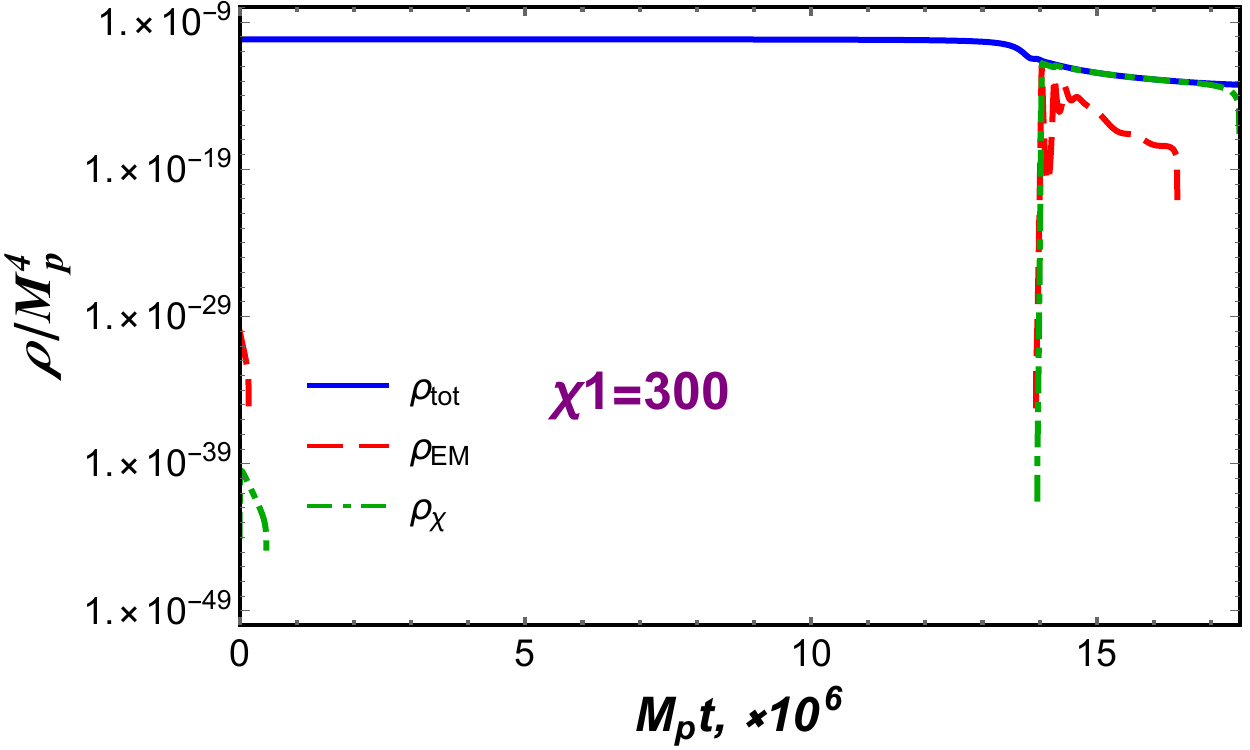}
		\caption{Panel a: using standard horizon scale $ k_{H}=aH $   and  panel b:  using effect of tachyonic instability by choosing $ K_{H}=aH|\zeta| $ .The time dependence (a)  of energy densities  for $ \chi_{1}=300 $ and  the simplest coupling function  Eq.\ref{Coupling-1} and  (b) for  the same coupling function  Eq.\ref{Coupling-1} and $ \chi_{1}=300 $ .In both panels there is no back-reaction problem . In each panel $ \rho_{tot} $ (blue), $ \rho_{EM}$ (red dashed line), $ \rho_{\chi} $ (green dashed line )  show the total energy density, electromagnetic energy density and energy density of charged particles due to  the Schwinger effect respectively.Note that in panel (a) we do not see effect of tachyonic instability because we use standard horizon scale  $ k_{H}=aH $ .In panel (b) we see that at very beginning of inflation both energy densities of electromagnetic field and produced charged particles due to the Schwinger effect vanish.This is the effect of tachyonic instability.Note that we increase the value of coupling constant and systematically see the same effect.}
		\label{Density-sim-2}
	\end{figure}

\subsection{Mode function}

Let us look at action (\ref{action-1}).In coulomb gauge $ A_{\mu}=\left(A_{0},A_{i}\right) $ with $ A_{i}=A^{T}_{i}+\partial_{i}\chi $ , where $ \partial_{i}A^{T}_{i}=0 $ ,we  have $\sqrt{-g}F_{\mu\nu}\tilde{F}^{\mu\nu}=4\epsilon_{ijk}{A^{T}_{i}}^{\prime}\partial_{j}A^{T}_{k} $ .Therefore,we find following equation

\begin{equation}
\label{Vector-Potential}
{A^{T}_{i}}^{\prime\prime}-\nabla^{2}A^{T}_{i}-I^{\prime}\left(\phi\right)\epsilon_{ijk}{A^{T}_{i}}^{\prime}\partial_{j}A^{T}_{k}=0
\end{equation}
The Fourier mode of equation (\ref{Vector-Potential}) is given by following relation
\begin{equation}
\label{Mode-1}
\mathcal {A}^{\prime\prime}_{\eta}+\left(k^{2}+hkI^{\prime}\left(\phi\right)\right)\mathcal{A}_{\eta}=0
\end{equation}
In terms of cosmic time , the above mode-function is given by following equation
\begin{equation}
\label{mode-3}
\ddot{\mathcal{A}_{h}}\left(t,k\right)+H\dot{\mathcal{A}_{h}}\left(t,k\right)+\left(\frac{k^{2}}{a^{2}\left(t\right)}+h\dot{I}\left(\phi\right)\frac{{k}}{a}\right)\mathcal{A}_{h}\left(t,k\right)=0
\end{equation}
In above equation  $ h=\pm $ shows the helicity.Also , the Fourier modes of the transverse part of vector potential can be written as  $  A^{T}_{i}\left(\eta,\textbf{k}\right)=\mathcal{A}_{+}\varepsilon_{+}+\mathcal{A}_{-}\varepsilon_{-} $.For more details about obtaining the above equation and decomposition function in Fourier  space and orthogonality relations, see our previous works in Refs.\cite{Kamarpour:2021,Kamarpour:G,Kamarpour:2022,Kamarpour:2023-I}.

\section{The Schwinger effect}
\label{sec-Schwinger}
We  only consider two expressions for  strong field regime.It has been shown  in Refs.\cite{Kamarpour:2022,Kamarpour:2023,Kamarpour:2023-I,Sobol-Gorbar:2021} that  the Schwinger effect in weak field regime is quite negligible and irrelevant to our calculations.Thus,for numerical calculations  we only need following  expressions.
\begin{equation}
\label{sigma}
\sigma_{s}=\frac{g_{s}}{12\pi^{3}}\frac{e^{3}E}{H}\exp\left({-\frac{\pi m^{2}}{|e E|}}\right), \hspace{1cm} s=b,f
\end{equation} 

In Eq.(\ref{sigma}) , $ g_{b}=1 $ and $ g_{f} $ are the number of spin degrees of freedom(d.o.f.).

The equation which explains produced charged particles from vacuum is:
\begin{equation}
\label{charged-e}
\dot{\rho}_{\chi}+4H\rho_{\chi}=2\rho_{E}\sigma_{s}
\end{equation}
In Eq.(\ref{charged-e}) $ 4H $ appears because we neglect mass and only consider massless charged particles.In fact, in comparison to Hubble parameter  the produced charged particles due to the Schwinger effect have smaller  mass.
\begin{figure}[ht]
	\centering
	\includegraphics[width=0.45\textwidth]{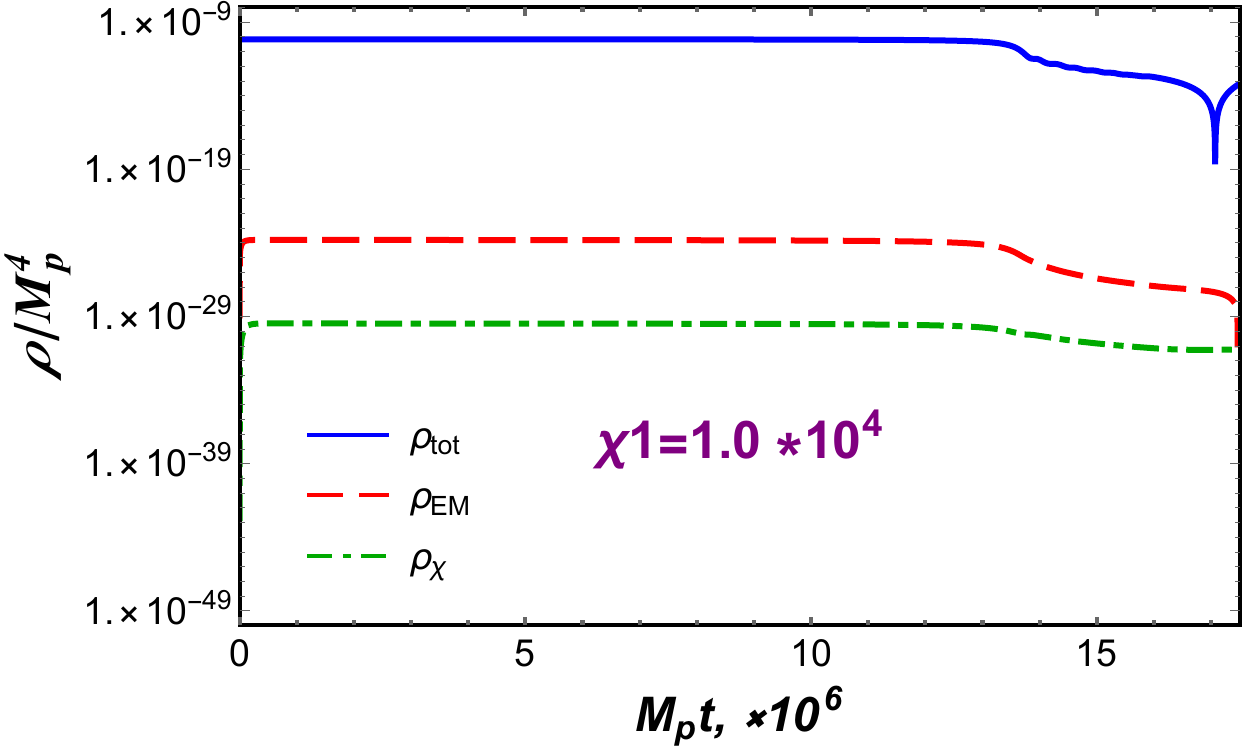}
	\hspace*{2mm}
	\includegraphics[width=0.45\textwidth]{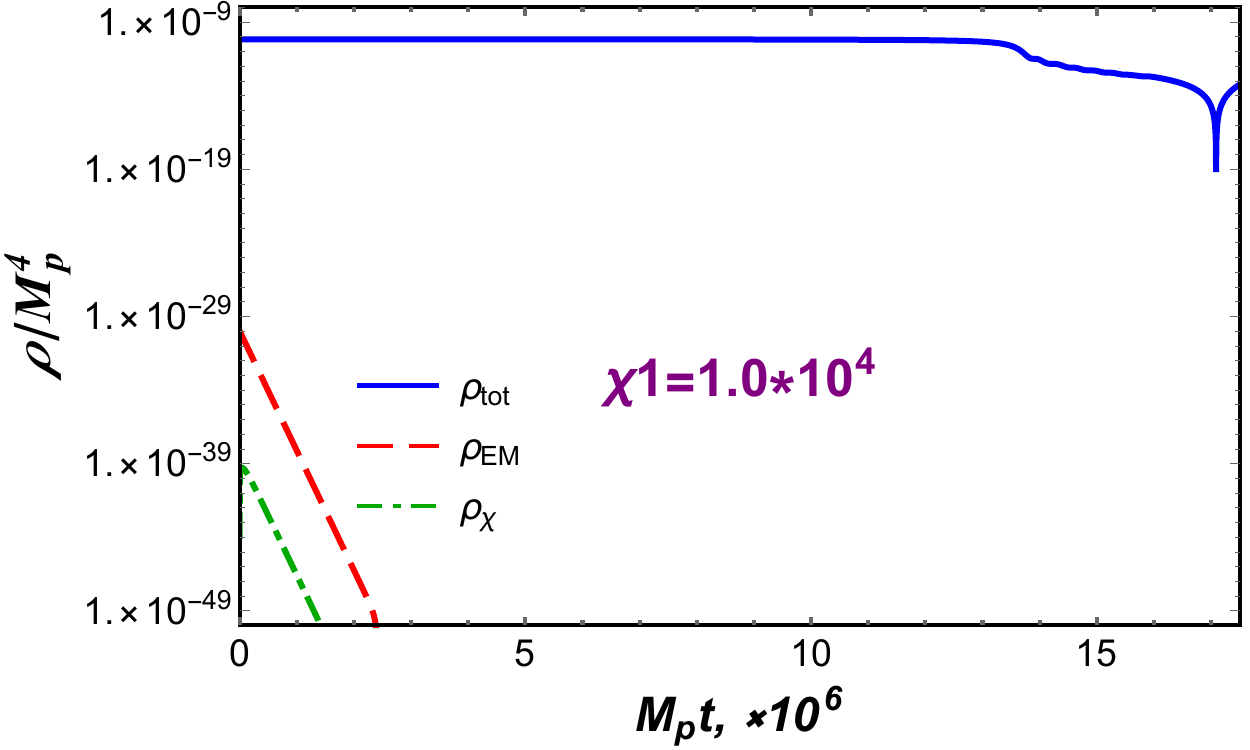}
	\caption{ Panel a: using standard horizon scale $ k_{H}=aH $   and  panel b:  using effect of tachyonic instability by choosing $ K_{H}=aH|\zeta| $ .The time dependence (a)  of energy densities  for $ \chi_{1}=1.0 \times 10^{4} $ and  non-minimal coupling function to gravity  Eq.\ref{Coupling-3} and  (b) for  the same coupling function  Eq.\ref{Coupling-3} and $ \chi_{1}=1.0 \times 10^{4} $ .In both panels there is no back-reaction problem . In each panel $ \rho_{tot} $ (blue), $ \rho_{EM}$ (red dashed line), $ \rho_{\chi} $ (green dashed line )  show the total energy density, electromagnetic energy density and energy density of charged particles due to  the Schwinger effect respectively.Note that in panel (a) we do not see  the effect of tachyonic instability because we use standard horizon scale  $ k_{H}=aH $ .In panel (b) we see that at the very beginning of inflation both energy densities of electromagnetic field and produced charged particles due to the Schwinger effect vanish.This is the effect of tachyonic instability.Note that  in both panels the Schwinger effect is quite negligible.}
	\label{Density-1}
\end{figure}

\begin{figure}[ht]
	\centering
	\includegraphics[width=0.45\textwidth]{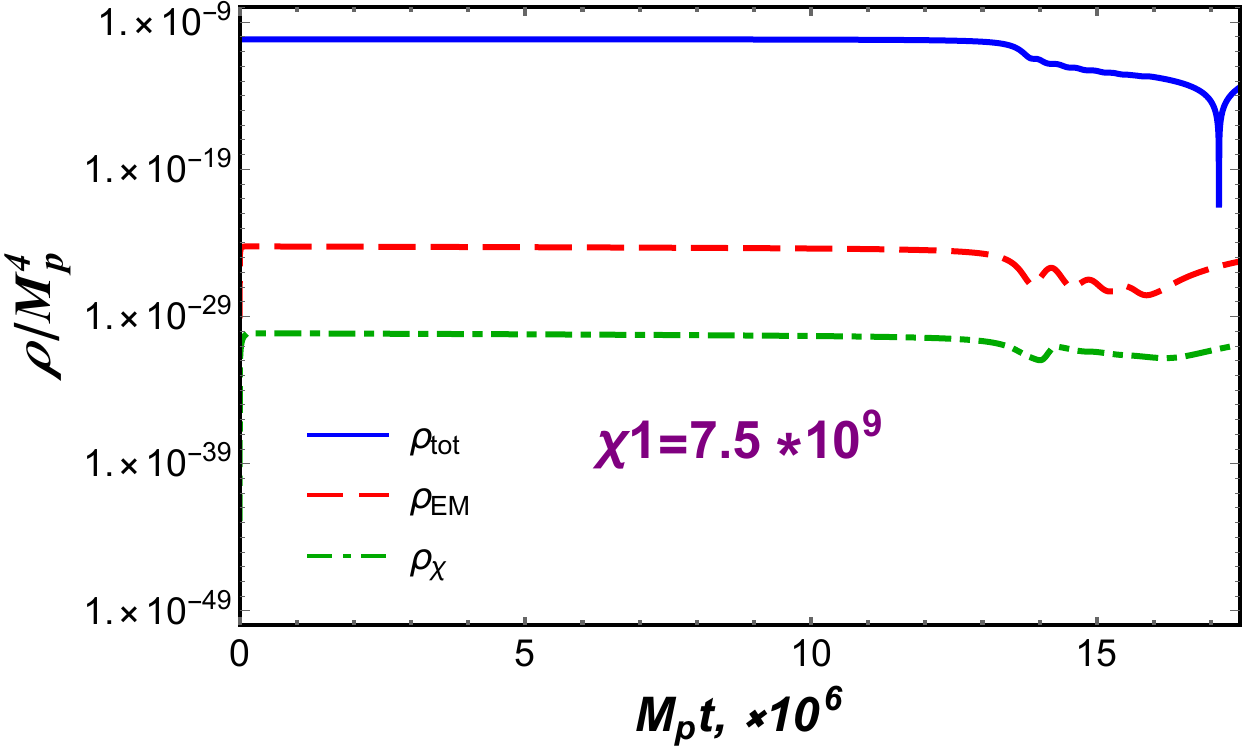}
	\hspace*{2mm}
	\includegraphics[width=0.45\textwidth]{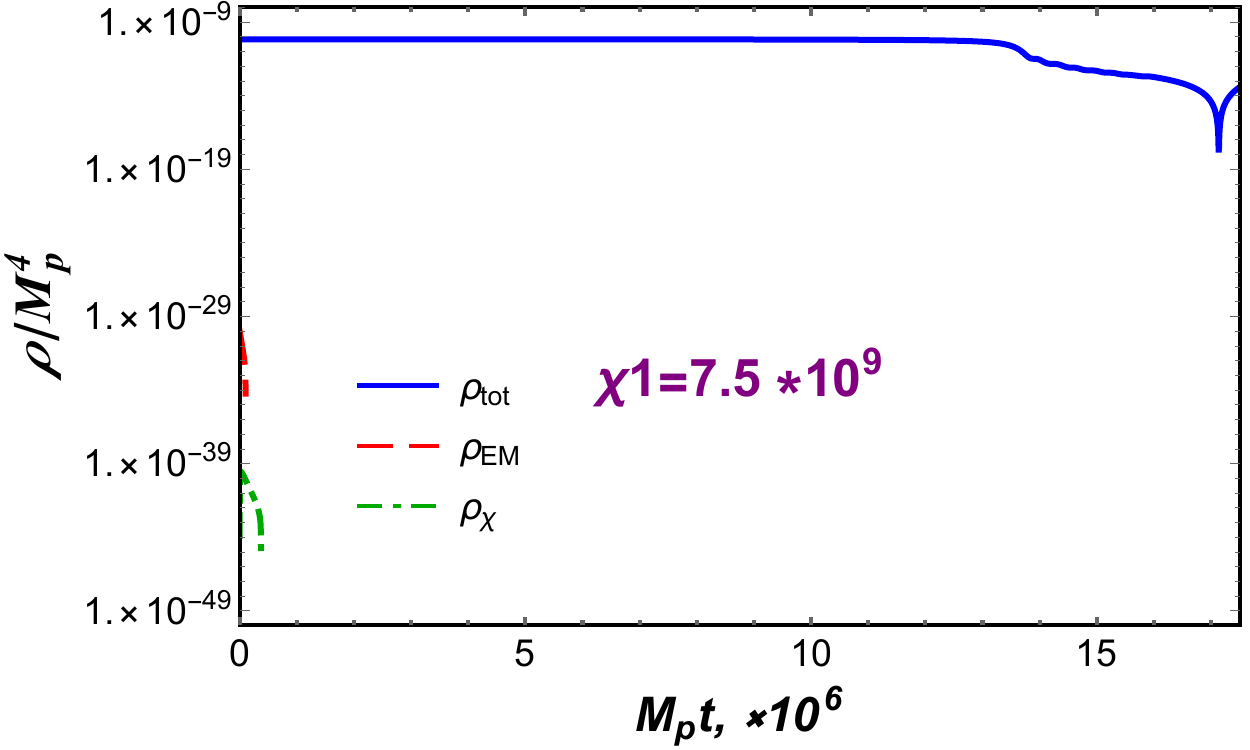}
	\caption{Panel a: using standard horizon scale $ k_{H}=aH $   and  panel b:  using effect of tachyonic instability by choosing $ K_{H}=aH|\zeta| $ .The time dependence (a)  of energy densities  for $ \chi_{1}=7.5\times 10^{9} $ and  non-minimal coupling function to gravity  Eq.\ref{Coupling-3} and  (b) for  the same coupling function  Eq.\ref{Coupling-3} and $ \chi_{1}=7.5 \times 10^{9} $ .In both panels there is no back-reaction problem . In each panel $ \rho_{tot} $ (blue), $ \rho_{EM}$ (red dashed line), $ \rho_{\chi} $ (green dashed line )  show the total energy density, electromagnetic energy density and energy density of charged particles due to  the Schwinger effect respectively.Note that in panel (a) we do not see  the effect of tachyonic instability because we use standard horizon scale  $ k_{H}=aH $ .In panel (b) we see that at very beginning of inflation both energy densities of electromagnetic field and produced charged particles due to the Schwinger effect vanish.This is the effect of tachyonic instability.Note that  in both panels the Schwinger effect is quite negligible.In addition  we increase the value of coupling constant and see systematic effect of tachyonic instability by vanishing energy densities at the very beginning of inflation.}
	\label{Density-3}
\end{figure}

\begin{figure}[ht]
\centering
	\includegraphics[width=0.45\textwidth]{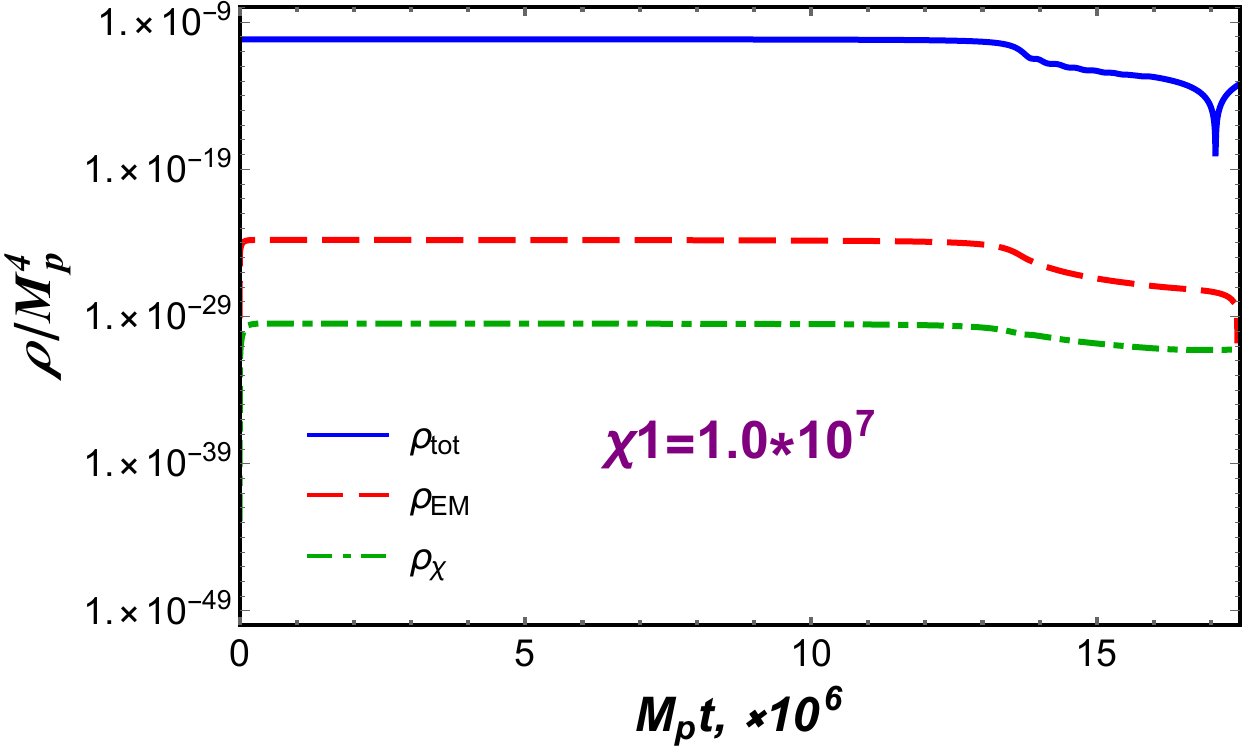}
	\hspace*{2mm}
	\includegraphics[width=0.45\textwidth]{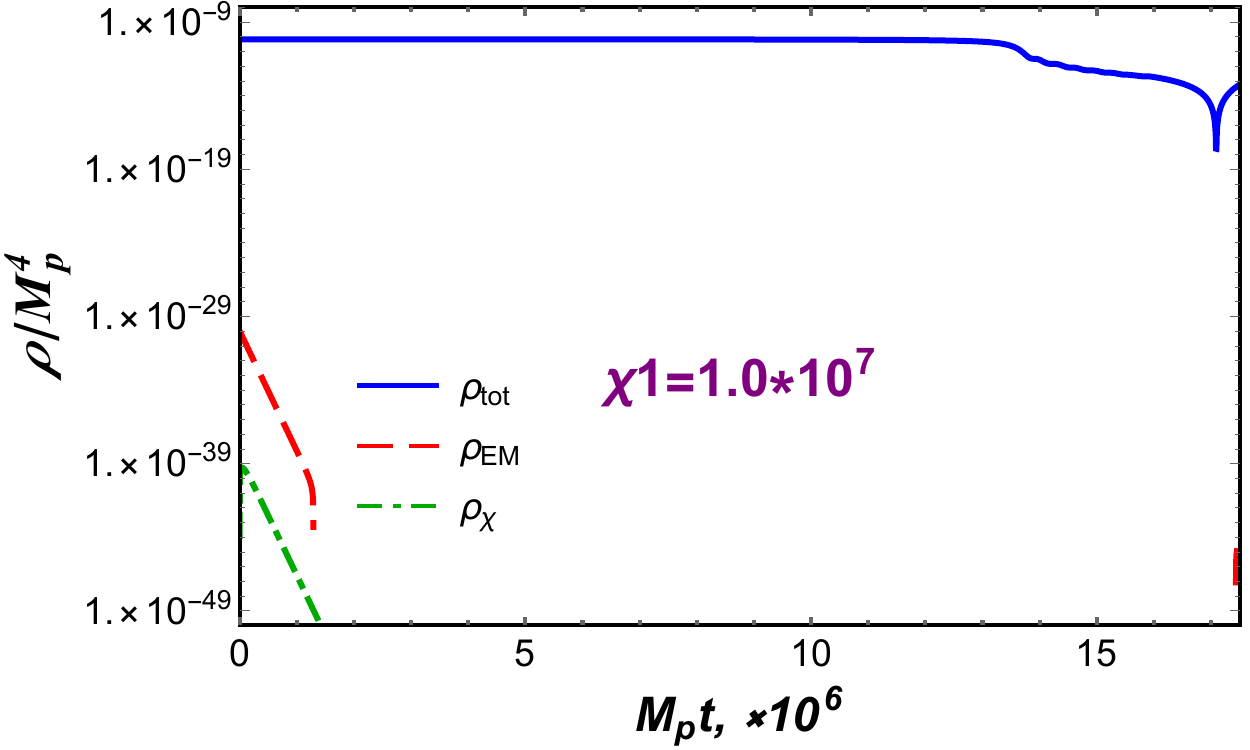}
	\caption{Panel a: using standard horizon scale $ k_{H}=aH $   and  panel b:  using effect of tachyonic instability by choosing $ K_{H}=aH|\zeta| $ .The time dependence (a)  of energy densities  for $ \chi_{1}=1.0 \times 10^{7} $ and  non-minimal coupling function to gravity  Eq.\ref{Coupling-3} and  (b) for  the same coupling function  Eq.\ref{Coupling-3} and $ \chi_{1}=1.0\times 10^{7} $ .In both panels there is no back-reaction problem . In each panel $ \rho_{tot} $ (blue), $ \rho_{EM}$ (red dashed line), $ \rho_{\chi} $ (green dashed line )  show the total energy density, electromagnetic energy density and energy density of charged particles due to  the Schwinger effect respectively.Note that in panel (a) we do not see  the effect of tachyonic instability because we use standard horizon scale  $ k_{H}=aH $ .In panel (b) we see that at very beginning of inflation both energy densities of electromagnetic field and produced charged particles due to the Schwinger effect vanish.This is the effect of tachyonic instability.Note that  in both panels the Schwinger effect is quite negligible.In addition,  we decrease the value of coupling constant and see systematic effect of tachyonic instability by vanishing energy densities at the very beginning of inflation. }
	\label{Density-4}
\end{figure}
\section{Numerical calculations}
\label{sec-Numerical}
\subsection{The simplest Coupling function}
For numerical calculations we should examine the consistency of CMB constraints and slow-roll parameters with anomalous scaling factor $ A_{I} $.The prediction for slow-roll parameters are consistent with the Planck data\cite{Planck:2015,Planck:2018} only for $ -4 <A_{I}<10 $.In numerical calculations we use $ n_{s}=0.960691 $ for spectral index and for tensor to scalar ratio $ r=0.00190286 $ and $ A_{I}=-3 $ ,number of e-folds $ N_{*}=60 $ and finally $ \phi_{0}=5.20568M_{p} $ , $ \phi_{e}=0.93347M_{p} $.See Refs.\cite{Kamarpour:2021,Kamarpour:G,Kamarpour:2022}.We label prefactor of potential (\ref{potential})  $ V_{0}= \frac{\lambda}{4}\frac{M_{p}^{4}}{\xi_{\eta}^{2}} $.We notice  that the value of $ V_{0}$ comes from the correct value of the scalar perturbations amplitude.

From Ref.\cite{Planck:2018} $ P_{s}=2.33361\times 10^{-9} $. For scalar perturbation amplitude of  $ P_{s}=2.33361\times 10^{-9} $ and $ \textbf{A}_{I}=-3 $ and $ \phi_{0}=5.20568M_{p} $ we obtain $ V_{0}=7.04914\times 10^{-11} $.See Ref.\cite{Kamarpour:2021} for more details.Below we introduce the simplest coupling function. 

%Before we start numerical calculations , it should be emphasized that required information about CMB constraints and slow-roll parameters such as spectral index,tensor to scalar ratio and other relevant informations $ \epsilon , \eta,n_{s} , r, $ are given in our previous works of Ref\cite{Kamarpour:2021,Kamarpour:G,Kamarpour:2022}.  For numerical calculations it is convenient to use the  following simplest coupling function.
\begin{equation}
\label{Coupling-1}
I\left(\phi\right)=\chi_{1}\frac{\phi}{M_{p}}
\end{equation} 
In above relation $ \chi_{1} $ is dimensionless coupling parameter.By using this coupling function we can estimate the values of coupling constant in numerical calculations.On the other hand this coupling function is very useful  for numerical calculations because gives us insight.Let us look at Eq.(\ref{Back-reaction}).If we use slow-roll condition then one may eliminate $ \ddot{\phi} $ and $ \dot{\phi} $ and find following relation
\begin{equation}
\frac{dV\left(\phi\right)}{d\phi}\sim -{I}^{\prime}\left(\phi\right)\textbf{E}\cdot\textbf{B}\sim-\frac{\chi_{1}}{M_{p}}\textbf{E}\cdot\textbf{B}
\end{equation}
 Using slow-roll parameter  $\epsilon=\frac{M_{p}^{2}}{2}\left(\frac{V^{\prime}}{V}\right)^{2}  $ and estimating in slow roll condition $ \rho_{inf}\sim V\left(\phi\right) $ holds , we find following approximation 
 \begin{equation}
 \label{epsilon}
\epsilon\sim\frac{1}{2}\left(\frac{\chi_{1}\textbf{E}\cdot\textbf{B}}{\rho_{inf}}\right)^{2}
 \end{equation}
 When we use approximation for $ \textbf{E}\cdot\textbf{B} $ then the above relation will be useful.See \cite{Sobol-Gorbar:2021}.
 
 \subsection{Non-minimal coupling to gravity}
 
We choose coupling function $ I\left(\phi\right) $ from conformal transformation $ \tilde{g}_{\mu\nu}=\Omega^{2}g_{\mu\nu} $ and achieve  following relation.See \cite{Kamarpour:2021,Kamarpour:G,Kamarpour:2022,Sobol:2021A}
\begin{equation}
\label{Coupling-2}
I\left(\phi\right)=12\chi_{1}e^{\left(\sqrt{\frac{2}{3}}\frac{\phi}{M_{p}}\right)}\left[\frac{1}{3M_{p}^{2}}\left(4V\left(\phi\right)\right)+\frac{\sqrt{2}}{\sqrt{3}M_{p}}\left(\frac{dV}{d\phi}\right)\right]
\end{equation}
In above equation the coupling constant $\chi_{1}$ has dimension $ M_{p}^{-2} $ .Inserting potential (\ref{potential})  into Eq.(\ref{Coupling-2}) the non-minimal  coupling function is given by following equation
\begin{equation}
\label{Coupling-3}
I\left(\phi\right)=\frac{16V_{0}}{M_{p}^{2}}\chi_{1}e^{\sqrt{\frac{2}{3}}\frac{\phi}{M_{p}}}\left(1-e^{-\sqrt{\frac{2}{3}}\frac{\phi}{M_{p}}}\right)\left[1+\frac{A_{I}}{32\pi^{2}}\ln\left(e^{\sqrt{\frac{2}{3}}\frac{\phi}{M_{p}}}-1\right)+\frac{A_{I}}{64\pi^{2}}\right]
\end{equation}
By taking derivative of Eq.(\ref{Coupling-3}) we find
\begin{equation}
\label{D-Coupling}
\dot{I}\left(\phi\right)=V_{0}\sqrt{\frac{2}{3}}\frac{1}{M^{3}_{p}}16\chi_{1}\dot{\phi}e^{\left(\sqrt{\frac{2}{3}}\frac{\phi}{M_{p}}\right)}\left[1+\frac{\textbf{A}_{I}}{32\pi^{2}}\ln\left(e^{\sqrt{\frac{2}{3}}\frac{\phi}{M_{p}}}-1\right)+\frac{3\textbf{A}_{I}}{64\pi^{2}}\right],
\end{equation}
Now we must switch on the Schwinger effect.For this , we 
 numerically solve equations (\ref{Back-reaction} , \ref{Friedmann-2} ,\ref{EM-Density} , \ref{charged-e} ) by using Eq.(\ref{sigma}) into Eq.(\ref{charged-e}) and setting $ \textbf{E}\cdot\textbf{B}\sim \sqrt{2\rho_{E}}\sqrt{2\rho_{B}}=2\sqrt{\rho_{E}\rho_{B}}\sim\rho_{EM} $.For this approximation we argue that in axial coupling $ \rho_{E}\sim \rho_{B} $  that we have already shown.See Refs.\cite{Figueroa:2018,Notari:2016,Fujita:2015,Kamarpour:2021,Kamarpour:2022,Kamarpour:2023-I}
\begin{figure}[ht]
	\centering
	\includegraphics[width=0.45\textwidth]{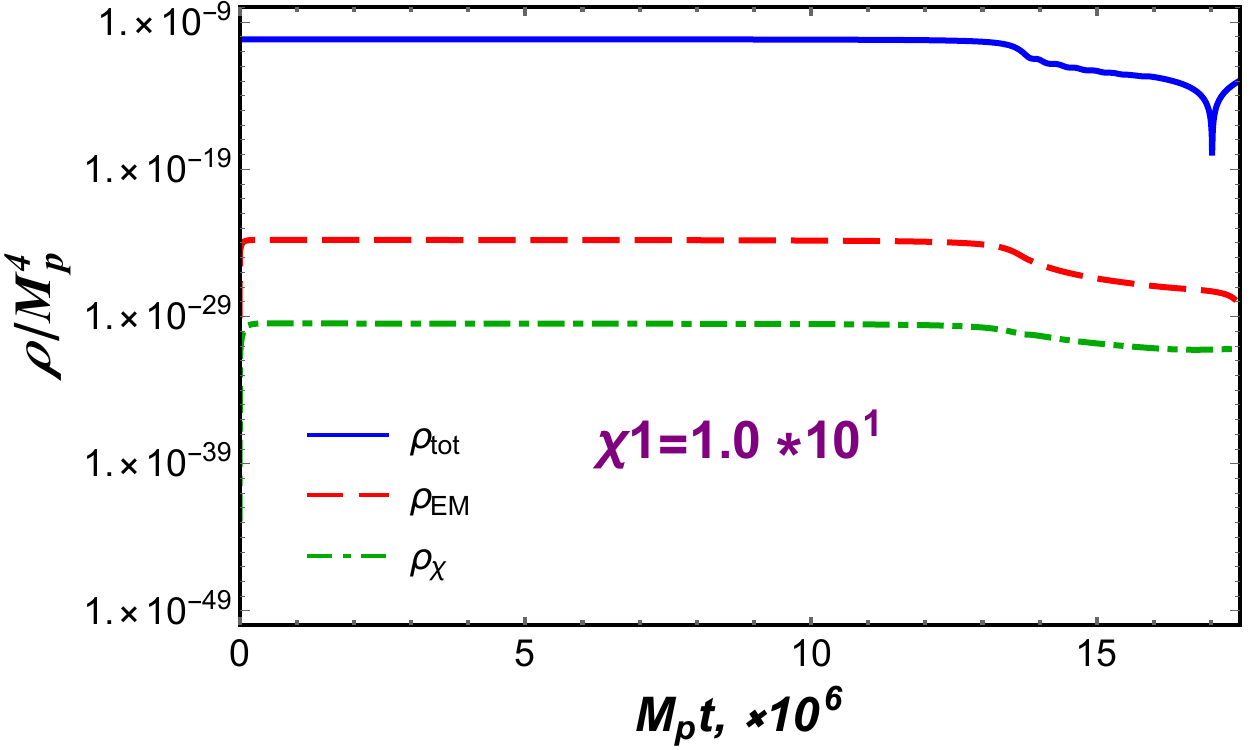}
	\hspace*{2mm}
	\includegraphics[width=0.45\textwidth]{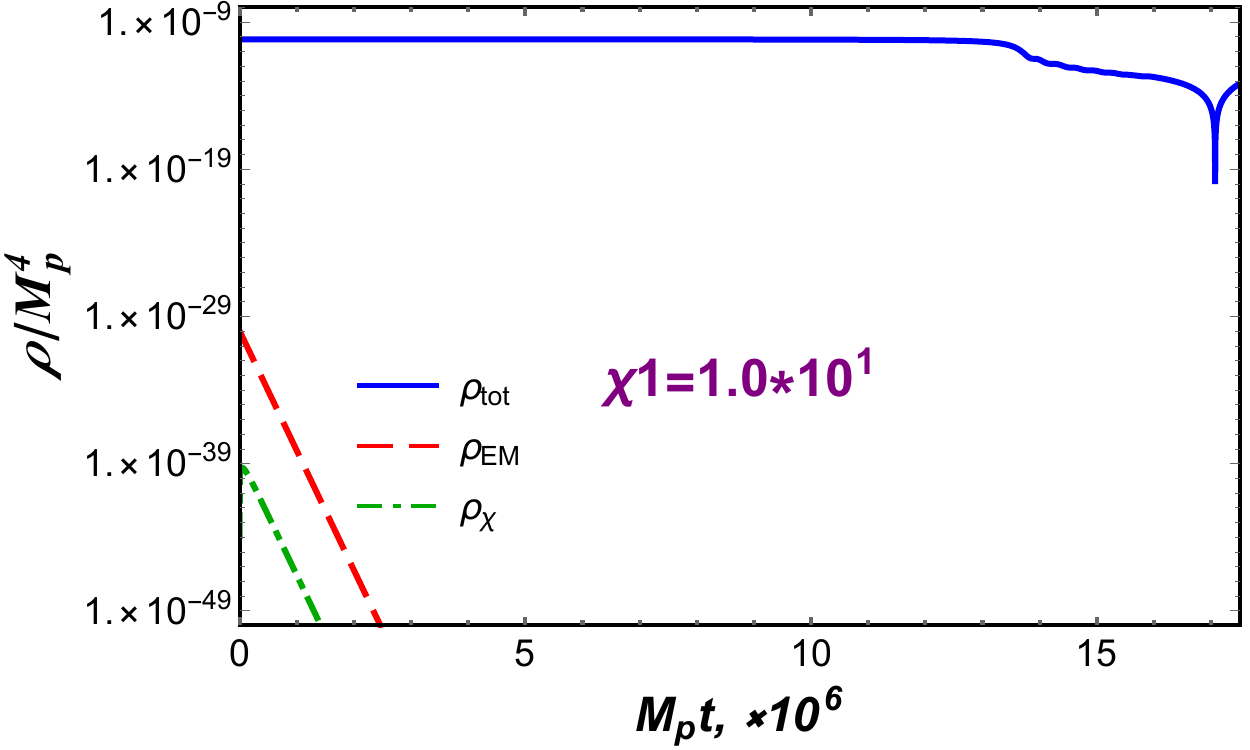}
	\caption{Panel a: using standard horizon scale $ k_{H}=aH $   and  panel b:  using effect of tachyonic instability by choosing $ K_{H}=aH|\zeta| $ .The time dependence (a)  of energy densities  for $ \chi_{1}=1.0 \times 10^{1} $ and  non-minimal coupling function to gravity  Eq.\ref{Coupling-3} and  (b) for  the same coupling function  Eq.\ref{Coupling-3} and $ \chi_{1}=1.0\times 10^{1} $ .In both panels there is no back-reaction problem . In each panel $ \rho_{tot} $ (blue), $ \rho_{EM}$ (red dashed line), $ \rho_{\chi} $ (green dashed line )  show the total energy density, electromagnetic energy density and energy density of charged particles due to  the Schwinger effect respectively.Note that in panel (a) we do not see  the effect of tachyonic instability because we use standard horizon scale  $ k_{H}=aH $ .In panel (b) we see that at the very beginning of inflation both energy densities of electromagnetic field and produced charged particles due to the Schwinger effect vanish.This is the effect of tachyonic instability.Note that  in both panels the Schwinger effect is quite negligible.In addition,  we decrease the value of coupling constant and choose the lowest value and see systematic effect of tachyonic instability by vanishing energy densities at the very beginning of inflation.}
	\label{Density-5}
\end{figure}
\subsection{Tachyonic instability}
We should add boundary term to the right hand side of equation (\ref{EM-Density}) but we need to discuss about tachyonic instability.Le us look at Eq.(\ref{mode-3}).We must set $ \ddot{\mathcal{A}_{h}}\left(t,k\right)=0 $ and $ \dot{\mathcal{A}_{h}}\left(t,k\right)=0 $. 

Term in bracket determines tachyonic instability.Tachyonic instability begins when $ h=- $ and $ \dot{I}\left(\phi\right)\frac{k}{a}\geq \frac{k^{2}}{a^{2}} $ or $ \frac{k}{aH}\leq\frac{{I}^{\prime}\left(\phi\right)\dot{\phi}}{H} $.We introduce the parameter $ \zeta $ so that
\begin{equation}
\label{zeta}
  \zeta=\frac{{I}^{\prime}\left(\phi\right)\dot{\phi}}{H} 
\end{equation}
Therefore, the condition for tachyonic instability is $ \frac{k}{aH}\leq|\zeta| $.The critical value for momentum is $ k_{c}=aH|\zeta| $.Thus allowed modes must satisfy $ k<k_{c} $ in order to be detectable.

The above discussion will lead  us to include the required boundary term to the right hand side of Eq.(\ref{EM-Density}).  
\subsection{Initial condition and boundary term}
Initial condition for solving equations (\ref{Mode-1}) or (\ref{mode-3}) was determined by Bunch-Davies.  
 
\begin{equation}
\mathcal{A}_{h}\left(t,k\right)=\frac{1}{\sqrt{2k}}e^{-ik\eta}, \hspace{1cm} k\eta\longrightarrow\infty
\end{equation}
Power spectrum of electric field is defined by\cite{Subramanian:2016,Durrer:2013}
\begin{equation}
\label{power-spectrum}
\frac{d\rho_{E}}{d\ln k}=\frac{k^{3}}{\left(2\pi\right)^{2}}\frac{1}{a^{2}}\left(|\frac{\partial\mathcal A_{+}(t,k)}{\partial t}|^{2}+|\frac{\partial\mathcal A_{-}(t,k)}{\partial t}|^{2}\right).
\end{equation} 
By using Eq.(\ref{power-spectrum}) we find required equation for boundary term in axial coupling\cite{Sobol:2018,Kamarpour:2022,Kamarpour:2023,Kamarpour:2023-I}.
\begin{equation}
\label{Last-1}
\left(\dot{\rho}_{E}\right)_{H}=\frac{d\rho_{E}}{dk}\arrowvert_{k=k_{H}}\cdot\frac{dk_{H}}{dt}=\frac{H^{5}}{8\pi^{2}},\hspace{.5cm}k_{H}=aH
\end{equation} 
 One writes Eq.(\ref{Last-1})by setting $ k_{H}=k_{c}=aH|\zeta| $ .Thus the required relation for boundary term in axial coupling  is given by\cite{Sobol-Gorbar:2021}.
\begin{equation}
\label{Last-2}
\left(\dot{\rho}_{E}\right)_{H}=\frac{d\rho_{E}}{dk}\arrowvert_{k=k_{H}}\cdot\frac{dk_{H}}{dt}=\frac{H^{5}|\zeta|^{3}}{\pi^{2}},\hspace{.5cm} k_{H}=k_{c}=aH|\zeta|
\end{equation} 
We must emphasize that the Eqs.(\ref{Last-1}) or (\ref{Last-2}) should be added to the right hand side of Eq.(\ref{EM-Density}) separately in order to investigate which one is more appropriate choice.
 
All remains to be done is to solve Eqs.(\ref{Friedmann-2} , \ref{Back-reaction} , \ref{EM-Density} , \ref{charged-e}) and to obtain required results.Before that let us look at figures.

Figures (\ref{Inflaton-1} ,  \ref{Inflaton-3}) show the time dependence  of inflation field without Schwinger effect (blue line )and  with the Schwinger effect (red dashed line)  for various values of  parameter $ \chi_{1} $ and non-minimal coupling function to gravity of equation (\ref{Coupling-3}) . Note that for all values of coupling parameter  $  \chi_{1} $  there is no back-reaction and  the Schwinger effect is quite negligible.These figures are compatible with  figures (\ref{Density-1},\ref{Density-3},\ref{Density-4},\ref{Density-5}).We see the energy densities of produced charged particles and electromagnetic filed  are negligible in panel (a) and in panel (b) the energy densities vanish due to effect of tachyonic instability.

Figures (\ref{Inflaton-4},\ref{Inflaton-5}) show  the time dependence of  inflaton field for the simplest coupling function  Eq.\ref{Coupling-1} (a) and  for the same coupling  Eq.\ref{Coupling-1} (b)  for the same values  of parameter $ \chi_{1} $.Both without(blue) the Schwinger effect and with (red)the Schwinger effect.We see small differences in oscillatory parts of two panels.In fact choosing tachyonic instability  $ K_{H}=aH|\zeta| $ has effect on inflaton field in oscillatory part.Note that , we increase the value of coupling parameter $ \chi_{1} $ and we see  that the effect of tachyoinc instability  in panel (b) is bigger in both figures.

Figures (\ref{Density-sim-1},\ref{Density-sim-2}, show the time dependence (a)  of energy densities  for $ \chi_{1}= 300,160 $ and  the simplest coupling function  Eq.\ref{Coupling-1} and  (b) for  the same coupling function  Eq.\ref{Coupling-3} and $ \chi_{1}= 300,160 $ .In both panels there is no back-reaction problem .

 In each panel $ \rho_{tot} $ (blue), $ \rho_{EM}$ (red dashed line), $ \rho_{\chi} $ (green dashed line )  show the total energy density, electromagnetic energy density and energy density of charged particles due to  the Schwinger effect respectively.Note that in panel (a) we do not see effect of tachyonic instability because we use standard horizon scale  $ k_{H}=aH $ .In panel (b) we see that at very beginning of inflation both energy densities of electromagnetic field and produced charged particles due to the Schwinger effect vanish.This is the effect of tachyonic instability.Note that  we increase  the value of coupling constant and systematically see the same effect.

Figures show  (\ref{Density-1},  \ref{Density-3}, \ref{Density-4},\ref{Density-5}) the time dependence (a)  of energy densities  for $ \chi_{1}=1.0 \times 10^{4},7.5\times 10^{9},1.0\times 10^{7}, 1.0 \times 10^{1} $ and  non-minimal coupling function to gravity  Eq.\ref{Coupling-3} and  (b) for  the same coupling function  Eq.\ref{Coupling-3} and the same values for $ \chi_{1} $ .

In both panels there is no back-reaction problem . In each panel $ \rho_{tot} $ (blue), $ \rho_{EM}$ (red dashed line), $ \rho_{\chi} $ (green dashed line )  show the total energy density, electromagnetic energy density and energy density of charged particles due to  the Schwinger effect respectively.Note that in panel (a) we do not see  the effect of tachyonic instability because we use the standard horizon scale  $ k_{H}=aH $ .In panel (b) we see that at very beginning of inflation both energy densities of electromagnetic field and produced charged particles due to the Schwinger effect vanish.This is the effect of tachyonic instability.Note that  in both panels the Schwinger effect is quite negligible.In addition,  we decrease the value of coupling constant and choose the lowest value and see systematic effect of tachyonic instability by vanishing energy densities at the very beginning of inflation.

\section{Conclusion}
\label{sec-conclusion}
In this work we examined the influence of tachyonic instability on  the Schwinger effect in Higgs inflation model by axial coupling.Our study was divided into two parts by using two boundary terms for identifying the horizon scale.

In first part , we assumed $ \rho_{E}\sim \rho_{B} $ and for this reason we only included $ \rho_{EM} $ in the Friedmann equation (\ref{Friedmann-2}).The validation of this assumption  was mentioned by considering axial coupling between electromagnetic field and inflaton field.In addition, the correctness of this  assumption was  confirmed in literature before \cite{Figueroa:2018,Notari:2016,Fujita:2015,Kamarpour:2021,Kamarpour:2022,Kamarpour:2023-I} .Also,in this part we use the standard horizon scale for boundary terms $ k_{H}=aH $.

We used two coupling functions , the simplest coupling of Eq.(\ref{Coupling-1}) and non-minimal coupling to gravity of Eq.(\ref{Coupling-3}).In first part ,we incorporated the Schwinger effect in our action and considered the equation of motion for the inflaton field , taking into account the back-reaction term $ \textbf{E}\cdot\textbf{B} $ in the right hand side of equation (\ref{Back-reaction}) and added a gauge invariant action to the  whole action \ref{action-1}.Then we finalized system of closed equations and performed numerical calculations by utilizing the coupling functions (\ref{Coupling-1} , \ref{Coupling-3})  and boundary term (\ref{Last-1}) in relations (\ref{Friedmann-2} , \ref{Back-reaction} , \ref{EM-Density} , \ref{charged-e}).

We plotted  figures (\ref{Inflaton-1}  , \ref{Inflaton-3} ,\ref{Inflaton-4}, \ref{Inflaton-5} ,\ref{Density-sim-1} , \ref{Density-sim-2} , \ref{Density-1} , \ref{Density-3}  , \ref{Density-4} , \ref{Density-5}) and  observed that when the horizon scale is $ k_{H}=aH $ , there is no back-reaction  and for non-minimal coupling function to gravity (\ref{Coupling-3}) the Schwinger effect is quite negligible.

Additionally and more importantly, when we use the horizon scale  $ k_{H}=aH $ and the simplest coupling function ((\ref{Coupling-1})) the Schwinger effect is considerable and plays important role in magneto-genesis.These results in first part are in accordance with our previous works in Refs \cite{Kamarpour:2022,Kamarpour:2023-I}.

In second part,we activated another boundary term in order to examine the influence of tachyonic instability on the Schwinger effect.Therefore,  we identified  the new horizon scale $ K_{H}=aH|\zeta| , \zeta=\frac{{I}^{\prime}\left(\phi\right)\dot{\phi}}{H} $ .This is the scale at which a given Fourier mode Eq.(\ref{mode-3}) begins to become tachyonically unstable.See equation(\ref{Last-2}).

Subsequently, we performed numerical calculations  and noticed that ,the effect of choosing tachyonic instability is vanishing the energy densities of charged particles and electromagnetic field at the very beginning of inflation.

In fact,choosing this scale  does not alter conclusions of the first part.The influence of this horizon scale is vanishing the energy densities of both electromagnetic field and produced charged particles at  the very beginning of inflation but does not alter final conclusion. 

Finally,in accordance  to our previous works in Refs.\cite{Kamarpour:2022,Kamarpour:2023-I} on Higgs inflation model in which both magneto-genesis for non-minimal function to gravity and the Schwinger effect  for the simplest coupling function were considerable,in this work we discovered only at the very beginning of inflation both energy densities of electromagnetic field and created charged particle vanish due to effect of  tachyoinc instability. 

%One may estimate due to existence of strong back-reaction problem , magneto-genesis by axial coupling in this model is impossible.But this needs separate investigation and should be addressed elsewhere. 

\begin{acknowledgments}
	
	The author would like to express  gratitude to S. Vilchinskii, E.V. Gorbar, and O. Sobol for their valuable insights and discussions during the preparation of this manuscript. Special thanks are also extended to O. Sobol for his assistance in creating the figures presented in this paper.

\end{acknowledgments}
 
\section*{Data Availability Statement}	

The Author confirms that this manuscript has no associated data in a data repository.

\section*{Declaration of competing interest }
The authors declare that they have no known competing financial interests or personal relationships
that could have appeared to influence the work reported in this paper.

%\bibliography{review}

\end{document}